\pdfoutput=1
\documentclass[10pt,a4paper]{article}
\usepackage[T2A]{fontenc}
\usepackage[utf8]{inputenc}
\usepackage[english]{babel}
\usepackage{amssymb,graphicx}
\usepackage{yfonts}
\usepackage{amsmath,amsfonts,mathtools}
\usepackage{amsthm}
\usepackage{framed}
\usepackage{fullpage}
\usepackage{enumitem} 
\usepackage[makeroom]{cancel}
\usepackage[export]{adjustbox} 
\usepackage{microtype}
\usepackage{hyperref}

\newcommand{\normord}[1]{:\mathrel{\mkern2mu #1 \mkern2mu}:}

 \newcommand{\tr}{\mathop{\mathrm{tr}}\nolimits}

\newcommand{\Ysf}{\mathsf{Y}}

\title{\textbf{On $R$-matrix formulation of $qq$-characters}}
\author{Mehmet Batu Bay\i nd\i rl\i$^{1,2}$\thanks{bayin004@umn.edu},
  Dilan Nur Demirta\c{s}$^{1,3}$\thanks{dilandemirtas@arizona.edu},
  Can Koz\c{c}az$^{1,4,5}$\thanks{can.kozcaz@boun.edu.tr}, Yegor
  Zenkevich$^{\text{6}}$\thanks{yegor.zenkevich@gmail.com}
  \footnote{On leave from ITMP MSU.}\\
  {$^1$\small\textit{Department of Physics, Bo\u{g}azi\c{c}i University, Istanbul, Turkey}}\\
  {$^2$\small\textit{School of Physics and Astronomy, University of Minnesota, Minneapolis, Minnesota 55455, USA}}\\
  {$^3$\small\textit{Department of Physics, University of Arizona,Tucson, AZ 85721, USA}}\\
  {$^4$\small\textit{Feza G\"{u}rsey Center for Physics and Mathematics, Bo\u{g}azi\c{c}i University, Istanbul, Turkey}}\\
  {$^5$\small\textit{Niels Bohr Institute, Copenhagen University, Blegdamsvej 17, Copenhagen, 2100, Denmark}}\\
  {$^6$\small\textit{Departments of Mathematics, University of
      California, Berkeley, USA}}} \date{}

\begin{document}

\maketitle

\begin{abstract}
  We introduce an $R$-matrix formulation of $qq$-characters and
  corresponding Frenkel-Reshetikhin deformed $W$-algebras. The
  $R$-matrix featuring in the construction is of Ding-Iohara-Miki
  (DIM) algebra, while the type of the $qq$-character is determined by
  the network of Fock representations corresponding to a web of
  5-branes geometrically engineering a quiver gauge theory. Our
  formulation gives a unified description of $qq$-characters of $A_n$
  type and their elliptic uplifts.
\end{abstract}

\tableofcontents

\section{Introduction}
\label{sec:introduction}

Quantum deformation is one of the central themes of modern ``physical
mathematics''. In particular, quantum deformations of Lie
algebras~\cite{Drinfeld, Jimbo} were discovered in connection with
lattice integrable models in statistical physics (see e.g.\ the
classic book~\cite{Baxter:1982zz}), but soon found applications in
many other areas such as knot theory~\cite{Witten:1988hf},
supersymmetric gauge
theories~\cite{Minahan:2006sk},~\cite{Nekrasov:2009ui}, and $2d$
conformal field theory
(CFT)~\cite{Bazhanov:1994ft}--\cite{Bazhanov:1998dq}.

To capture the intricacies of the representation theory of quantum
affine Lie algebras $U_q(\widehat{\mathfrak{g}})$ Knight~\cite{Knight}
and Frenkel-Reshetikhin~\cite{FR} have introduced the
notion of $q$-characters which can be thought of as generalizations of
conventional characters with more parameters. The principal
ingredient in the construction of $q$-character is the $R$-matrix of
$U_q(\widehat{\mathfrak{g}})$. The $q$-characters are partial traces
of the $R$-matrix taken in finite-dimensional representations, so that
algebraic properties of the $R$-matrix imply identities for the
$q$-characters.

It also turned out that there is a \emph{second} quantum deformation
of $q$-characters involving parameter $t$, called $qq$-characters,
natural from both geometric~\cite{Nak} and algebraic~\cite{FR}
points of view. The deformation turns the parameters in the character
into non-commutative operators so that the characters themselves
generate a nontrivial algebra --- the $(q,t)$-deformed $W$-algebra,
$W_{q,t}(\mathfrak{g})$.

It would be desirable to have an $R$-matrix construction of
$qq$-characters similar to that of $q$-characters. Some attempts along
these lines have been made in~\cite{Liu}, but that paper dealt
exclusively with a specific (and complicated) case of $\mathfrak{g} =
\widehat{\mathfrak{gl}}_1$. In the current paper we propose a more
general $R$-matrix formalism for $qq$-characters, which to our
knowledge has not been suggested previously.

There are some parallels between our formalism and that for
$q$-characters: we also use an $R$-matrix as a central tool. However,
the details differ a lot. Firstly, we use one and the same ``master''
$R$-matrix to obtain all the $qq$-characters for the root systems
$\mathfrak{g} = A_n$. The $R$-matrix is that of the Ding-Iohara-Miki
(DIM)~\cite{DIM} or quantum toroidal algebra
$\mathcal{A}=U_{q,t}(\widehat{\widehat{\mathfrak{gl}}}_1)$. In this
sense both the root system $\mathfrak{g}$ \emph{and} the
representation featuring in the $qq$-character are free parameters
that can be varied in our approach without changing the underlying
algebraic structure.

The second important difference between our formalism
and~\cite{FR} is that we don't take a trace of our $R$-matrix,
but instead a (partial) matrix element thereof. The states in the
matrix element are what encodes the representation featuring in the
$qq$-character.

The technical framework that we employ to derive the expressions for
$qq$-characters is refined topological string. This allows us to
interpret various algebraic expressions as refined amplitudes on
certain toric Calabi-Yau (CY) three-folds. In fact the toric diagram
of the three-fold can be viewed as a lattice statistical
model~\cite{Awata:2016mxc} with crossings corresponding to DIM
$R$-matrices~\cite{Z20}.

Let us note that ours is not the first attempt to obtain
$qq$-characters in the refined topological string
framework. In~\cite{MMZ} (see also \cite{Bourgine:2017jsi})
$qq$-characters were identified with specific generators of the
algebra $\mathcal{A}$ inserted between combinations of refined
topological vertices. However, the generators were not viewed as
coming from an $R$-matrix. In~\cite{Kimura:2017auj} $qq$-characters
were engineered by certain linear combinations of Lagrangian brane
insertions, but no algebraic interpretation was given. In contrast, in
the present paper we provide an explicit algorithm to draw a toric
diagram whose refined partition function produces a given
$qq$-character.

In the remaining part of the introduction we recall briefly the
$R$-matrix construction of $q$-characters of quantum affine Lie
algebras~\cite{FR} (sec.~\ref{sec:q-characters-quantum}), and then
introduce the general idea of our approach to $qq$-characters and
their connection with Type IIB 5-branes and refined topological
strings (sec.~\ref{sec:qq-char-quant}). In this way we demonstrate
both the similarities as well as differences between the two
approaches before delving into full technical description.

The rest of the paper is organized as follows. In
sec.~\ref{sec:dim-r-matrices} we introduce the DIM $R$-matrix which
can be interpreted as a spectator brane insertion. In
sec.~\ref{sec:stack-cross-top} we show how the $R$-matrix formulas
give rise to the coproduct of the generating current of the DIM
algebra and match it with some previous works on the relation between
DIM algebra and $qq$-characters. In sec.~\ref{sec:norm-order-comm} we
show how $qq$-characters are obtained from the $R$-matrix insertion
using Wick's theorem. We apply our formalism to fundamental $A_n$ type
$qq$-characters in sec.~\ref{sec:fund-qq-char} and to higher
representations of $A_n$ in sec.~\ref{sec:higher-qq-characters}. In
sec.~\ref{sec:elliptic-lift-qq} we consider the uplift of the
formalism to elliptic $A_n$ $qq$-characters.  We present our
conclusions in sec.~\ref{sec:conclusions}.

\subsection{$q$-characters and quantum affine algebras}
\label{sec:q-characters-quantum}

$q$-characters are partial traces of an $R$-matrix. Let us be more
precise. We consider finite-dimensional representations of quantum
affine algebra $U_q(\widehat{\mathfrak{g}})$. They have trivial
central charges. The universal $R$-matrix $\mathcal{R} \in
U_q(\widehat{\mathfrak{g}}) \otimes U_q(\widehat{\mathfrak{g}})$ has
the schematic form~\cite{KhT}
\begin{equation}
  \label{eq:4}
  \mathcal{R} \sim \mathcal{R}_{+} \mathcal{R}_0 \mathcal{R}_{-},
\end{equation}
where the parts $\mathcal{R}_{\pm}$ will be irrelevant for us. The
``diagonal'' part $\mathcal{R}_0$ has the form\footnote{In order to
  eliminate cumbersome coefficients in the formulas we use the
  normalization of the generators which differs from the more standard
  one from~\cite{FR}.}
\begin{equation}
  \label{eq:5}
  \mathcal{R}_0 = \exp \left[ \sum_{n \geq 1} \sum_{i\in I} h_{i,n} \otimes \tilde{h}_{i,-n}  \right],
\end{equation}
where the index $i \in I$ runs over the set of simple roots of
$\mathfrak{g}$ and $h_{i,n}$ and $\tilde{h}_{i,n}$ are certain special
bases in the commutative subalgebra $U_q(\widehat{\mathfrak{h}})
\subset U_q(\widehat{\mathfrak{g}})$. The $q$-character of a
representation $V$ has the schematic form (we gloss over some
technical details such as prefactors)
\begin{equation}
  \label{eq:6}
  \chi_q(V) \sim P (\tr_V \otimes 1) \mathcal{R} \sim (\tr_V \otimes 1) \mathcal{R}_0,
\end{equation}
where $P$ is a certain projection operator whose job is effectively to
eliminate $\mathcal{R}_{\pm}$ factors. The untraced generators
$h_{i,n}$ commute and can be viewed as parameters of the
$q$-character.

A more compact way to pack the parameters is to introduce their
generating functions $Y_{i,a}$ (where $a \in \mathbb{C}^{\times}$),
which look like
\begin{equation}
  \label{eq:7}
  Y_{i,a}  \sim \exp \left( \sum_{n \geq 1} a^n \tilde{h}_{n,i} \right). 
\end{equation}

Since $h_{i,n}$ form a commuting set of operators, they can be
simultaneously diagonalized and one can study their joint
eigenspectra. The theorem proven in~\cite{FR} is that in
finite-dimensional representations of $U_q(\widehat{\mathfrak{g}})$
the eigenvalues have the form
\begin{equation}
  \label{eq:8}
  h_{i,n}|v\rangle \sim \left[ \sum_{r=1}^{k_i} a_{i,r}^n -
    \sum_{s=1}^{l_i} b_{i,s}^n \right] |v\rangle,
\end{equation}
where $a_{i,r}$, $b_{i,s}$ are complex numbers and $k_i$, $l_i$ are
finite integers. Moreover, for the highest weight vector there are no
$b_{i,s}$'s, i.e.\ $l_i=0$.

Plugging Eq.~\eqref{eq:8} into Eq.~\eqref{eq:6} and comparing with
Eq.~\eqref{eq:7} we find that the $q$-character is a Laurent
polynomial in $Y_{i,a}$, and that the highest weight of a
finite-dimensional representation corresponds to a monomial with only
positive powers of $Y_{i,a}$:
\begin{equation}
  \label{eq:9}
  \chi_q(V) = \sum_{|v\rangle \in V} c_v \prod_{i,r} Y_{i_r,a_r}
  \prod_{j,s} Y^{-1}_{j_s,b_s} = \prod_{i,r} Y_{i_r,a_r} + \ldots
\end{equation}

Properties of the $R$-matrix guarantee that $q$-characters are
additive and multiplicative. The definition~\eqref{eq:6} of the
character can be drawn as a picture:
\begin{equation}
  \label{eq:10}
  \chi_q(V)\quad =\quad \includegraphics[valign=c]{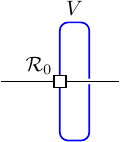}
\end{equation}
In Eq.~\eqref{eq:10} the square on the intersection of the blue and
black lines represents the $R$-matrix of
$U_q(\widehat{\mathfrak{g}})$, the blue line is the representation $V$
over which one takes a trace, and the horizontal black line is the
space in which the commuting $Y_{i,a}$ operators act.

\subsection{$qq$-characters and DIM algebra}
\label{sec:qq-char-quant}

In $qq$-characters the $Y_{a,i}$ operators are no longer
commuting. Instead they become vertex operators acting on (a tensor
power of) the Fock space, the Hilbert space of a free boson.

To engineer a $qq$-character we start with the universal $R$-matrix of
the DIM algebra and evaluate it in the tensor product of the vertical
and horizontal Fock spaces, which we denote by
$\mathcal{F}_{q,t^{-1}}^{(0,1)}(w)$ and
$\mathcal{F}_{q,t^{-1}}^{(1,0)}(u)$ respectively (see
Appendix~\ref{sec:basic-facts-about} for the definitions of the
algebra and representations). The $R$-matrix turns out to be equal to
the refined topological string amplitude on degenerate resolved
conifold~\cite{Z20}:
\begin{equation}
  \label{eq:11}
  \mathcal{R}|_{\mathcal{F}_{q,t^{-1}}^{(0,1)}(w)\otimes
    \mathcal{F}_{q,t^{-1}}^{(1,0)}(u)}
  =\quad  \includegraphics[valign=c]{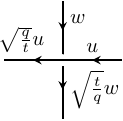} \quad =\quad     \includegraphics[valign=c]{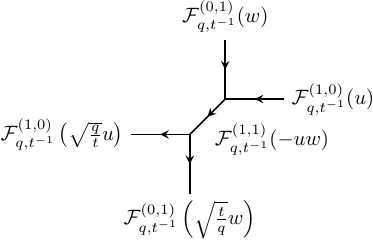}.
\end{equation}
The lines in Eq.~\eqref{eq:11} have two different physical
interpretations related by dualities in string theory.
\begin{enumerate}
\item The lines may be viewed as forming a toric diagram of a toric
  Calabi-Yau (CY) three-fold, on which refined topological strings
  propagate. For example the second picture in Eq.~\eqref{eq:11}
  represents the resoved conifold (the total space of the bundle
  $\mathcal{O}(-1) \oplus \mathcal{O}(-1) \to \mathbb{P}^1$). The Fock
  space associated with a leg of the diagram is the space of states of
  refined topological string wrapping the corresponding two-cycle
  inside the CY. Triple junctions in the picture are refined
  topological vertices~\cite{RefTopVert}. The lines passing on top of
  each other without intersecting as in the first picture in
  Eq.~\eqref{eq:11} correspond to deformed conifold geometry. The
  second equality in Eq.~\eqref{eq:11} expresses the geometric
  transition between deformed and resolved conifold in refined
  topological string theory~\cite{Gopakumar:1998ki}.

\item Alternatively, the lines may be thought of as $(p,q)$ 5-branes
  in Type IIB string theory on a flat background\footnote{The equivalence between these two interpretations is shown in \cite{Leung:1997tw}.}. In this
  interpretation the Fock space on a leg of the diagram is the Hilbert
  spaces of BPS states bound to the corresponding 5-brane. The triple
  junctions in the picture are junctions of 5-branes. Two lines
  passing on top of each other without intersecting are simply branes
  lying in different two-dimensional planes. In gauge theories living
  on the worldvolumes of the branes this corresponds to passing to the
  Higgs branch.
\end{enumerate}
We will use the terms lines and branes interchangeably henceforth.

Degenerate crossing Eq.~\eqref{eq:11} has special
properties. Specifically, its matrix elements along the vertical Fock
representation satisfy the following selection rule:
\begin{equation}
  \label{eq:12}
\boxed{  \includegraphics[valign=c]{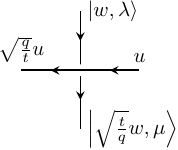} \quad
  =\quad 0 \quad \text{if} \quad \mu \nsubseteq \lambda,}
\end{equation}
where $\lambda$ and $\mu$ are Young diagrams labelling the states in
the Fock space.

To get the $qq$-characters of $A_{n-1}$ type corresponding to the
first fundamental representation $V_{\omega_1} \simeq \mathbb{C}^n$ we
draw the toric diagram involving $n$ horizontal lines and a vertical
``spectator'' brane crossing the horizontal lines. The external states
on the vertical line are $|\square\rangle$ and
$|\varnothing\rangle$. For example for $n=3$:
\begin{equation}
  \label{eq:13}
  \chi_{qq}^{A_2}(\mathbb{C}^3) =  \quad      \includegraphics[valign=c]{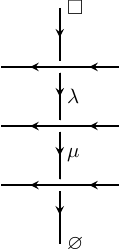}.
\end{equation}
The external states are indicated above and below the vertical
spectator brane. Every crossing in Eq.~\eqref{eq:13} is an $R$-matrix
insertion and a sum over intermediate states $\lambda$ and $\mu$ is
assumed. According to the selection rule Eq.~\eqref{eq:12} there are only
three possibilities for $\lambda$ and $\mu$, namely $(\lambda,\mu) =
(\varnothing, \varnothing)$, $(\square, \varnothing)$, and
$(\square,\square)$. As we will see in sec.~\ref{sec:norm-order-comm}
each of these possibilities produces a distinct combination of
$Y$-operators acting in the tensor product of three horizontal Fock
spaces. The resulting three terms reproduce the $qq$-character of the
defining representation $\mathbb{C}^3$ of $A_2$.

To get $qq$-characters corresponding to higher representations one
needs to insert several spectator branes. We explain the procedure in
detail in sec.~\ref{sec:higher-qq-characters}.

Notice how our approach is similar in spirit (the main object is the
$R$-matrix), but also markedly different from
sec.~\ref{sec:q-characters-quantum}:
\begin{enumerate}
\item The $R$-matrices belong to a larger algebra (DIM) with two
  quantum deformation parameters instead of one.

\item The same $R$-matrix is used to build the $qq$-characters of all
  $A_n$ types\footnote{We expect that all classical and affine series
    can in fact be treated in our framework.}.

\item Instead of a trace one needs to take a matrix element of the
  $R$-matrix.
    
\item The representations of quantum affine Lie algebras are encoded
  in the external states living in the vertical Fock representations
  of DIM.
\end{enumerate}

One could insert the spectator brane~\eqref{eq:13} into a given toric
diagram to get an average of the corresponding $qq$-character. Some
toric diagrams correspond to $5d$ $\mathcal{N}=1$ supersymmetric gauge
theories, in which the $qq$-characters form a special class of
observables called $\mathcal{X}$-observables as explained
in~\cite{Nekrasov:2017gzb}. For example, the diagram
\begin{equation}
  \label{eq:1}
    \includegraphics[valign=c]{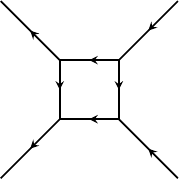}
\end{equation}
corresponds to a pure $SU(2)$ gauge theory. Due to the properties of
the $R$-matrix, the spectator brane can be inserted anywhere in the
picture, for example
\begin{equation}
  \label{eq:2}
    \includegraphics[valign=c]{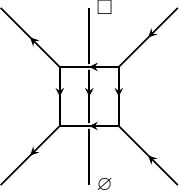}
\end{equation}
We will demonstrate the appearance of the $\mathcal{X}$-observables
from pictures such as Eq.~\eqref{eq:2} in sec.~\ref{sec:norm-order-comm}.

\section{DIM $R$-matrices as brane crossings}
\label{sec:dim-r-matrices}

In this section we give a more technical description of DIM
$R$-matrices and provide an explicit formula for the crossing operator
introduced in Eq.~\eqref{eq:11}. According to~\cite{Z20}, the
$R$-matrix for the tensor product of vertical and horizontal Fock
spaces can be written as a composition of two intertwining operators,
or refined topological vertices (see
Appendix~\ref{sec:basic-facts-about} for definitions of the Fock
representations).

The formulas for the intertwiners were obtained in~\cite{AFS}. Up to
prefactors they are given by the following vertex operators (see
Eqs.~\eqref{eq:84}, \eqref{eq:85} in
Appendix~\ref{sec:algebr-engin-su2} for complete expressions):
\begin{multline}
  \label{eq:35}
\Psi^{\lambda}(x) =
\quad \includegraphics[valign=c]{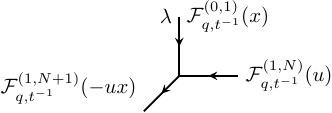} \quad  \\
\sim \,\normord{\exp \left[ \sum_{n \neq 0} \frac{x^{-n}}{n}
\left( \frac{1}{1-q^{-n}} - (1-t^n) \mathrm{Ch}_{\lambda}(q^{-n},
  t^{-n})\right) a_n\right]},
\end{multline}
\begin{multline}
\Psi^{*}_{\mu}(y) =
\quad \includegraphics[valign=c]{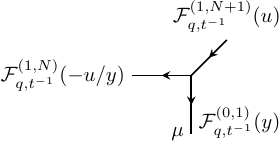} \\
\sim \, \normord{ \exp \left[ -\sum_{n \neq 0} \frac{y^{-n}}{n} \left( \frac{t}{q} \right)^{\frac{|n|}{2}}
\left( \frac{1}{1-q^{-n}} - (1-t^n) \mathrm{Ch}_{\mu}(q^{-n},
  t^{-n})\right) a_n\right]},\label{eq:55}
\end{multline}
where
\begin{equation}
   \mathrm{Ch}_{\lambda}(q,t) = \sum_{(i,j)\in \lambda} q^{j-1}
  t^{1-i}, \label{eq:36}
\end{equation}
and the free boson modes $a_n$ satisfy the $(q,t)$-deformed Heisenberg
commutation relations:
\begin{equation}
  [a_n, a_m] = n \frac{1 - q^{|n|}}{1-t^{|n|}}\delta_{n+m,0}. \label{eq:86}
\end{equation}

Combining the intertwiners Eqs.~\eqref{eq:35},~\eqref{eq:55} as drawn in
Eq.~\eqref{eq:11} we get the (the matrix element of) the $R$-matrix~\cite{Z22}:
\begin{multline}
  \label{eq:3}
  \mathcal{R}^{\lambda}_{\mu}(w,u,N) \stackrel{\mathrm{def}}{=}
  \Psi^{\lambda}(w) \Psi_{\mu}\left( \sqrt{\frac{t}{q}}w \right)
  =\includegraphics[valign=c]{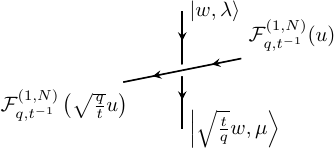}
  \quad \\
  = (-w)^{-N|\lambda|}u^{|\lambda|} \left( - \sqrt{\frac{t}{q}} w
  \right)^{N|\mu|} \left( \frac{q}{u} \sqrt{\frac{t}{q}}
  \right)^{|\mu|} N_{\lambda \mu}\left( \frac{q}{t} \right)
  \frac{q^{n(\lambda^{\mathrm{T}})+ n(\mu^{\mathrm{T}})}}{f_{\lambda}
    c_{\lambda} c_{\mu}} \left( \frac{f_{\mu}}{f_{\lambda}} \right)^N   \\
  \times :\! \exp \left[ - \sum_{n \geq 1} \frac{w^n}{n} \frac{\left(
        1-
        \left( t/q \right)^n\right)}{(1-q^n)} a_{-n} \right] \times\\
  \times \exp \left[ - \sum_{n \neq 0} \frac{(1-t^n) }{n} \left(
      \mathrm{Ch}_{\lambda}(q^{-n},t^{-n}) - \left( \frac{t}{q}
      \right)^{\frac{|n|-n}{2}} \mathrm{Ch}_{\mu}(q^{-n},t^{-n})
    \right)w^{-n} a_n \right]\! :
\end{multline}
where $N_{\lambda \mu}(x)$ is given by Eq.~\eqref{eq:14} and
\begin{align}
  n(\lambda^{\mathrm{T}}) &= \sum_{(i,j)\in \lambda} (j-1),\label{eq:113}\\
  c_{\lambda} &= \prod_{(i,j)\in \lambda} \left( 1 - q^{\lambda_i - j}
    t^{\lambda_j^{\mathrm{T}} - i +1} \right),\label{eq:114}\\
  f_{\lambda} &= \prod_{(i,j)\in \lambda} \left( - q^{j-\frac{1}{2}}
    t^{\frac{1}{2} - i} \right)=(-1)^{|\lambda|}q^{\frac{\|
      \lambda\|^2}{2}}t^{-\frac{\|\lambda^{\mathrm{T}}\|^2}{2}}=(-1)^{|\lambda|}
  q^{n(\lambda^{\mathrm{T}})+|\lambda|/2} t^{-n(\lambda)-|\lambda|/2},\label{eq:89}\\
  \|\lambda\|^2 &= \sum_{i=1}^{l(\lambda)} \lambda_i^2 =
  \sum_{(i,j)\in \lambda} (2j-1) = 2n(\lambda^{\mathrm{T}})
  +|\lambda|.\label{eq:115}
\end{align}

Several remarks are in order:
\begin{enumerate}
\item Eq.~\eqref{eq:3} is a generalization of Eq.~\eqref{eq:11} in
  which the horizontal line ($(1,0)$ Fock representation in
  Eq.~\eqref{eq:11}) has a nontrivial slope $(1,N)$, $N \in
  \mathbb{Z}$.

\item DIM algebra has an infinite number of inequivalent coproducts
  $\Delta^{(s)}$ labelled by an irrational slope $s$ in the
  $\mathbb{Z}^2$ lattice. Each of the coproducts gives rise to its own
  set of intertwining operators and $R$-matrices. In all of the
  formulas in this paper the choice of DIM coproduct (also known as
  the preferred direction in refined topological strings) is
  understood to be vertical, $s = \infty$.
  
\item The exponent in the third line of Eq.~\eqref{eq:3} was obtained
  in~\cite{Z20} as the \emph{vacuum} matrix element of the
  $R$-matrix. The rest of the formula is responsible for its
  generalization to arbitrary states on the vertical brane.

\item $\mathcal{R}^{\lambda}_{\mu}$ satisfies the selection
  rule~\eqref{eq:12} due to the vanishing of the Nekrasov factor
  $N_{\lambda \mu} (q/t)$ in the second line of Eq.~\eqref{eq:3}.

\item The additional terms that appear in
  $\mathcal{R}^{\lambda}_{\mu}$ for nontrivial $\lambda$ and $\mu$
  have the form of a product of vertex operators sitting at points
  corresponding to boxes of $\lambda$ and $\mu$, respectively.

\item We will sometimes omit the $u$ and $N$ arguments in
  $\mathcal{R}^{\lambda}_{\mu}(w,u,N)$ when there is no possibility
  for confusion.
\end{enumerate}

The operator $\mathcal{R}^{\lambda}_{\mu}$ looks complicated, however
in our present study we will need to evaluate it only for $\lambda$, $ \mu$ equal
to $\square$ or $\varnothing$. Due to the selection rule Eq.~\eqref{eq:12}
there are three cases of this type:
\begin{enumerate}
\item $\lambda = \varnothing$, $\mu = \varnothing$:
  \begin{equation}
    \label{eq:15}
    \mathcal{R}^{\varnothing}_{\varnothing} (w,u,N) =  \exp \left[ - \sum_{n \geq 1} \frac{w^n}{n} \frac{\left( 1-
          \left( t/q \right)^n\right)}{(1-q^n)} a_{-n} \right]
  \end{equation}
  It is important that the exponent in Eq.~\eqref{eq:15} contains only
  negative bosonic modes, so it acts trivially on the bra vacuum state
  $\langle \varnothing|$.
  
\item $\lambda = \square$, $\mu = \varnothing$:
  \begin{equation}
    \label{eq:16}
    \mathcal{R}^{\square}_{\varnothing} (w,u,N) =
  -\sqrt{\frac{t}{q}} \frac{1 - \frac{q}{t}}{1 - t}  \mathcal{R}^{\varnothing}_{\varnothing} (w,u,N)\, x^{+}(w)|_{\mathcal{F}^{(1,N)}(u)},
  \end{equation}
  where $x^{+}(w)|_{\mathcal{F}^{(1,N)}(u)}$ is the generating current
  of the DIM algebra in the $(1,N)$ Fock representation given by
  Eq.~\eqref{eq:63} in Appendix~\ref{sec:representations}. Notice that
  the product $\mathcal{R}^{\varnothing}_{\varnothing}(w)\, x^{+}(w)$
  is automatically normal ordered. We will use this fact when we
  compute correlators in sec.~\ref{sec:norm-order-comm}.

\item $\lambda = \square$, $\mu = \square$:
  \begin{equation}
    \label{eq:18}
    \mathcal{R}^{\square}_{\square} (w,u,N) = 
    \frac{1-q}{1-t} \psi^{-}\left( \left( t/q
      \right)^{\frac{1}{4}} w \right)|_{\mathcal{F}^{(1,N)}(u)}
    \mathcal{R}^{\varnothing}_{\varnothing} (w,u,N),
  \end{equation}
  where $\psi^{-}(w)$ is another generating current of the DIM
  algebra given by Eq.~\eqref{eq:66} in
  Appendix~\ref{sec:representations}. The product in Eq.~\eqref{eq:18}
  is also automatically normal ordered.
\end{enumerate}

Having the basic ingredients --- the
crossings~\eqref{eq:15},~\eqref{eq:16} and~\eqref{eq:18} --- we can
proceed to combine them. Naturally, there are two ways:
\begin{enumerate}
\item One can stack crossings on top of each other, so that a single
  vertical line crosses a number of horizontal ones. This will produce
  the $qq$-characters of the first fundamental representation
  $\mathbb{C}^n$ of $A_{n-1}$ which we investigate in
  sec.~\ref{sec:stack-cross-top}.

\item The crossings can be joined along the horizontal legs. This will
  produce $qq$-characters of higher representations as described in
  sec.~\ref{sec:higher-qq-characters}.
\end{enumerate}

\section{DIM coproduct and fundamental $qq$-characters from brane
  crossings}
\label{sec:stack-cross-top}

Consider a vertical spectator brane intersecting $n$ horizontal ones
as shown in Eq.~\eqref{eq:13} with external state $|\square\rangle$ on
top and $\langle \varnothing |$ at the bottom. Every brane
intersection gives rise to the $R$-matrix operator Eq.~\eqref{eq:3}. We
claim that such an intersection produces the $qq$-character of the
defining representation $\mathbb{C}^n$ of $A_{n-1}$ algebra.

To simplify our presentation in this section we restrict ourselves to
the case when the branes intersecting the vertical brane are strictly
horizontal, i.e.\ correspond to Fock representations of type
$(1,0)$. However, the same arguments work for $(1,N)$ lines too. Let
us follow the example of Eq.~\eqref{eq:13}, where there are three
horizontal lines and two diagrams on the intermediate edges $\lambda$
and $\mu$.

As we have mentioned in sec.~\ref{sec:qq-char-quant}, selection rules
severely restrict the set of Young diagrams on the intermediate
vertical edges. There are three possible pairs $(\lambda,\mu)$
contributing to the answer. Let us write down the operators
corresponding to each of these pairs:
\begin{enumerate}
\item $(\lambda,\mu) = (\varnothing, \varnothing)$:
  \begin{multline}
    \label{eq:23}
    \includegraphics[valign=c]{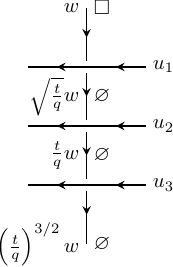} \quad =
    \quad \mathcal{R}^{\square}_{\varnothing}(w,u_1,0) \otimes
    \mathcal{R}^{\varnothing}_{\varnothing}\left(
      \sqrt{\frac{t}{q}}w,u_2,0\right) \otimes
    \mathcal{R}^{\varnothing}_{\varnothing}\left(
      \frac{t}{q}w,u_1,0\right)
  \end{multline}
  which can alternatively be expressed using Eq.~\eqref{eq:16} as
  \begin{align}\nonumber
   &-\sqrt{\frac{t}{q}} \frac{1 - \frac{q}{t}}{1-t}
    \mathcal{R}^{\varnothing}_{\varnothing}(w,u_1,0) \otimes
    \mathcal{R}^{\varnothing}_{\varnothing}\left(
      \sqrt{\frac{t}{q}}w,u_2,0\right) \otimes
    \mathcal{R}^{\varnothing}_{\varnothing}\left(
      \frac{t}{q}w,u_1,0\right) (x^{+}(w)|_{\mathcal{F}(u_1)} \otimes
    1 \otimes 1 )=\\
&= \mathcal{T}^{\varnothing}_{\varnothing}(w) (x^{+}(w)|_{\mathcal{F}(u_1)} \otimes
    1 \otimes 1 )   = u_1 \mathcal{T}^{\varnothing}_{\varnothing}(w) \normord{\exp \left[ -
      \sum_{n \neq 0} \frac{1-t^n}{n} w^{-n} a_n^{(1)} \right]},
\end{align}
where $a^{(i)}_n$, $i=1,2,3$ are three sets of bosonic operators
acting on three horizontal lines and the operator $x^{+}(w)$ is
defined in Eq.~\eqref{eq:63}. The ``empty crossing'' part is given
by
\begin{equation}
  \label{eq:24}
  \mathcal{T}^{\varnothing}_{\varnothing}(w) = -\sqrt{\frac{t}{q}} \frac{1 - \frac{q}{t}}{1-t} \exp \left[ - \sum_{n \geq 1} \frac{w^n}{n} \frac{ 1-
        \left( t/q \right)^n}{1-q^n} \bar{a}_{-n} \right].
\end{equation}
where we have defined the ``diagonal part'' of the bosonic modes
\begin{equation}
  \label{eq:17}
  \bar{a}_{-n} = a_{-n}^{(1)} +
      \left( \frac{t}{q} \right)^{\frac{n}{2}} a_{-n}^{(2)} + \left(
        \frac{t}{q} \right)^n a_{-n}^{(3)}.
\end{equation}
Notice that $\mathcal{T}^{\varnothing}_{\varnothing}(w)$ has a natural
algebraic meaning of its own: it corresponds to the crossing with a
vertical spectator brane, but with empty external states on it. In
that case the selection rule~\eqref{eq:12} dictates that the Young
diagrams on the intermediate edges should all be empty. Therefore the
operator $\mathcal{T}^{\varnothing}_{\varnothing}(w)$ factorizes into
a product of operators each acting on its own horizontal brane (see
sec.~4.3 of~\cite{Z20}).
  
\item $(\lambda,\mu) = (\square, \varnothing)$:
  \begin{multline}
    \label{eq:25}
    \includegraphics[valign=c]{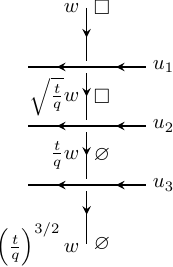} \quad =
    \quad b_{\square}(q,t)\, \mathcal{R}^{\square}_{\square}(w,u_1,0)
    \otimes \mathcal{R}^{\square}_{\varnothing}\left(
      \sqrt{\frac{t}{q}}w,u_2,0\right) \otimes
    \mathcal{R}^{\varnothing}_{\varnothing}\left(
      \frac{t}{q}w,u_1,0\right),
  \end{multline}
  which can be similarly expressed in terms of the operator
  $\mathcal{T}^{\varnothing}_{\varnothing}(w)$ and another operator
  acting on the tensor product of three horizontal Fock spaces,
  \begin{multline}
    \text{Eq. }\eqref{eq:25}=b_{\square}(q,t)\mathcal{T}^{\varnothing}_{\varnothing}(w)
    \left(\left.\psi^{-}\left( \left( t/q \right)^{\frac{1}{4}} w
        \right)\right|_{\mathcal{F}^{(1,0)}(u_1)}
      \otimes \left. x^{+} \left( \sqrt{\frac{t}{q}} w \right)\right|_{\mathcal{F}^{(1,0)}(u_2)} \otimes 1\right)=\\
    =b_{\square}(q,t)u_2 \mathcal{T}^{\varnothing}_{\varnothing}(w) \normord{\exp
      \left[ \sum_{n \geq 1} \frac{1-t^{-n}}{n} \left( 1 - \left(
            \frac{t}{q} \right)^n \right) w^n a_{-n}^{(1)} - \sum_{n
          \neq 0} \frac{1-t^n}{n} w^{-n} \left( \frac{q}{t}
        \right)^{\frac{n}{2}} a_n^{(2)} \right]},
  \end{multline}
  where
  \begin{equation}
    \label{eq:56}
    b_{\mu}(q,t)  = \langle P_\mu | P_\mu\rangle^{-1} =
    \prod_{(i,j)\in \lambda} \frac{ 1 - q^{\lambda_i - j}
      t^{\lambda_j^{\mathrm{T}} - i +1} }{ 1 - q^{\lambda_i - j+1}
      t^{\lambda_j^{\mathrm{T}} - i} } = \frac{c_{\lambda}}{c'_{\lambda}}
  \end{equation}
  is the inverse of the norm of Macdonald polynomial corresponding to
  the intermediate state on the vertical brane ($b_{\square} = \frac{1-t}{1-q}$) and $\psi^{-}(w)$ is
  defined in Eq.~\eqref{eq:66}.

 \item $(\lambda,\mu) = (\square, \square)$:
   \begin{multline}
     \label{eq:26}
         \includegraphics[valign=c]{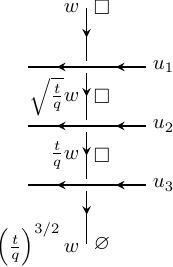} \quad =
    \quad b_{\square}(q,t)^2 \,
     \mathcal{R}^{\square}_{\square}(w,u_1,0) \otimes
    \mathcal{R}^{\square}_{\square}\left(
      \sqrt{\frac{t}{q}}w,u_2,0\right) \otimes
    \mathcal{R}^{\square}_{\varnothing}\left(
      \frac{t}{q}w,u_1,0\right),
\end{multline}
which can be rewritten as
\begin{multline}
b_{\square}^2 \mathcal{T}^{\varnothing}_{\varnothing}(w)
  \left(\left.\psi^{-}\left( \left( t/q \right)^{\frac{1}{4}} w
      \right)\right|_{\mathcal{F}^{(1,0)}(u_1)} \otimes
    \left.\psi^{-}\left( \left( t/q \right)^{\frac{3}{4}} w
      \right)\right|_{\mathcal{F}^{(1,0)}(u_2)}
    \otimes \left. x^{+} \left( \frac{t}{q} w \right)\right|_{\mathcal{F}^{(1,0)}(u_3)} \right)=\\
  =  b_{\square}^2u_3 \mathcal{T}^{\varnothing}_{\varnothing}(w) \normord{\exp
    \left[ \sum_{n \geq 1} \frac{1-t^{-n}}{n} \left( 1 - \left(
          \frac{t}{q} \right)^n \right) w^n \left( a_{-n}^{(1)} +
        \left( \frac{t}{q} \right)^{\frac{n}{2}} a_{-n}^{(2)} \right)
      - \sum_{n \neq 0} \frac{1-t^n}{n} w^{-n} \left( \frac{q}{t}
      \right)^n a_n^{(3)} \right]}.
   \end{multline}
\end{enumerate}

Collecting the terms for each pair of intermediate diagrams
$(\lambda,\mu)$ we get the following result for the crossing:
\begin{multline}
  \label{eq:27}
  \includegraphics[valign=c]{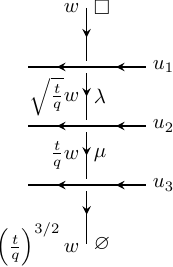} \quad = \quad
  \mathcal{T}^{\varnothing}_{\varnothing}(w) \Biggl\{
  x^{+}(w)|_{\mathcal{F}(u_1)} \otimes
  1 \otimes 1 +\\ +\left.\psi^{-}\left( \left( t/q \right)^{\frac{1}{4}} w
    \right)\right|_{\mathcal{F}^{(1,0)}(u_1)} \otimes \left. x^{+}
    \left( \sqrt{\frac{t}{q}} w
    \right)\right|_{\mathcal{F}^{(1,0)}(u_2)} \otimes 1+ \\
  + \left.\psi^{-}\left( \left( t/q \right)^{\frac{1}{4}} w
    \right)\right|_{\mathcal{F}^{(1,0)}(u_1)} \otimes
  \left.\psi^{-}\left( \left( t/q \right)^{\frac{3}{4}} w
    \right)\right|_{\mathcal{F}^{(1,0)}(u_2)} \otimes \left. x^{+}
    \left( \frac{t}{q} w
    \right)\right|_{\mathcal{F}^{(1,0)}(u_3)} \Biggr\}=\\
=\boxed{\mathcal{T}^{\varnothing}_{\varnothing}(w) \left. (\Delta \otimes 1)\Delta
    (x^{+}(w)) \right|_{\mathcal{F}^{(1,0)}(u_1)\otimes
    \mathcal{F}^{(1,0)}(u_2)\otimes \mathcal{F}^{(1,0)}(u_3)}}
\end{multline}
where in the last line we have used the formula for the coproduct of
the generating current $x^{+}(w)$ which can be found in
Eq.~\eqref{eq:58} in Appendix~\ref{sec:coproducts}.

At this point we would like to use some of the results of~\cite{MMZ}
summarized in Appendix~\ref{sec:qw_n-reduction}. It was shown there
that the insertion of the current $x^{+}(w)$ of the algebra
$\mathcal{A}$ acting in the tensor product of $m$ horizontal Fock
representations $\mathcal{F}^{(1,0)}_{q,t^{-1}}(u_i)$ generates the
fundamental $qq$-character of $A_{m-1}$ type. In Eq.~\eqref{eq:27} we
get almost the same result from the vertical brane insertion, but with
an additional operator
$\mathcal{T}^{\varnothing}_{\varnothing}(w)$. This factor, however,
will not affect most of the formulas that we get for the following
reason. We will consider the insertion of the
$R$-matrix/$qq$-character into some ``background'' network of
intertwining operators (refined topological vertices). As we will see
in sec.~\ref{sec:norm-order-comm} the extra operator
$\mathcal{T}^{\varnothing}_{\varnothing}(w)$ has very simple
commutation relations with refined topological vertices placed
elsewhere in the diagram. This will allow us to effectively get rid of
it in any given $qq$-character calculation by moving
$\mathcal{T}^{\varnothing}_{\varnothing}(w)$ to the very left of the
diagram, where it annihilates the bra vacuum state. Thus, we will
dismiss the $\mathcal{T}^{\varnothing}_{\varnothing}(w)$ in most of
the formulas.

To understand why the coproduct $(\Delta\otimes 1) \Delta (x^{+}(w))$ appears in
Eq.~\eqref{eq:27} we have to recall the Khoroshkin-Tolstoy
formula~\cite{KhT} for the universal $\mathcal{R}$-matrix. The
formula has the form:
\begin{equation}
  \label{eq:30}
  \mathcal{R} = P \mathcal{R}_0 \mathcal{R}_1 \mathcal{R}_2,
\end{equation}
where $P$ is the permutation operator exchanging the two factors in
the tensor product. The other three factors are
\begin{align}
  \mathcal{R}_0 &= e^{-c_1 \otimes d_1 - d_1 \otimes c_1 - c_2 \otimes d_2 - d_2 \otimes c_2},\label{eq:21}\\
  \mathcal{R}_1 &= \exp \left[ -\sum_{n \geq 1} \frac{\kappa_n}{n}
    e_{(0,n)} \otimes
      e_{(0,-n)}\right],\\
    \mathcal{R}_2&=
    1 + \kappa_1 \oint
    \frac{dz}{z} x^{-}(z) \otimes x^{+}(z) + \ldots\label{eq:22}
\end{align}
Let us explain the notations in Eqs.~\eqref{eq:21}--\eqref{eq:22} (see
Appendix~\ref{sec:basic-facts-about} for basic definitions related to
the DIM algebra):
\begin{enumerate}
\item $e_{(n,m)}$ are generators of the DIM algebra,

\item $c_1$, $c_2$ are the two central charges of the algebra,

\item $d_1$, $d_2$ are the grading operators,

\item $\kappa_n = (1-q^n)(1-t^{-n})(1-(t/q)^n)$, 

\item $x^{\pm}(z) = \sum_{n\in \mathbb{Z}} e_{(\pm 1,n)} z^{\mp n}$
  are DIM generating currents.

\item In Eq.~\eqref{eq:22} the terms replaced by the ellipsis involve
  DIM generators $e_{(n,m)}$ with $|n| \geq 2$. As we will see
  momentarily these terms will not contribute to the crossing
  operator~\eqref{eq:27} and therefore to the $qq$-character.
\end{enumerate}

The universal $R$-matrix~\eqref{eq:30} satisfies the fundamental
identity
\begin{equation}
  (\Delta\otimes 1)\mathcal{R} = \mathcal{R}_{12} \mathcal{R}_{13},\label{eq:20}
\end{equation}
where $\Delta$ is the coproduct given by
Eqs.~\eqref{eq:60}--\eqref{eq:61} and $\mathcal{R}_{12}$ (resp.\
$\mathcal{R}_{13}$) represents $\mathcal{R}$ acting on the first two
(res.\ first and third) factors in the triple tensor product.

It is the property in Eq.~\eqref{eq:20} that explains the appearance
of the coproduct of $x^{+}(w)$ in Eq.~\eqref{eq:27}. Indeed, consider
$\mathcal{R}$ acting in the tensor product of two representations: one
is a vertical Fock space $\mathcal{F}^{(0,1)}_{q,t^{-1}}(w)$ and the
other is itself a tensor product of three horizontal Fock spaces
$\mathcal{F}^{(1,0)}_{q,t^{-1}}(u_1)\otimes
\mathcal{F}^{(1,0)}_{q,t^{-1}}(u_2)
\otimes\mathcal{F}^{(1,0)}_{q,t^{-1}}(u_3)$ (see
Appendix~\ref{sec:basic-facts-about} for the definitions of the
relevant Fock representations). Due to the identity~\eqref{eq:20} the
$R$-matrix acting on the tensor product is
\begin{multline}
  \label{eq:28}
  (1 \otimes 1 \otimes \Delta) (1 \otimes \Delta) \mathcal{R}|_{\mathcal{F}^{(0,1)}(w) \otimes \mathcal{F}^{(1,0)}(u_1) \otimes
    \mathcal{F}^{(1,0)}(u_2) \otimes \mathcal{F}^{(1,0)}(u_3)}=\\
  = \mathcal{R}|_{\mathcal{F}^{(0,1)}(w) \otimes
    \mathcal{F}^{(1,0)}(u_1)}
  \mathcal{R}|_{\mathcal{F}^{(0,1)}(\sqrt{t/q} w) \otimes \mathcal{F}^{(1,0)}(u_2)} \mathcal{R}|_{\mathcal{F}^{(0,1)}(t/q w) \otimes \mathcal{F}^{(1,0)}(u_3)}.
\end{multline}

Let us compute the matrix element of the two sides of
Eq.~\eqref{eq:28} between the states $\langle \varnothing|$ and
$|\square\rangle$ in the vertical Fock representation. The r.h.s.\
gives the diagram in Eq.~\eqref{eq:27} while the l.h.s.\ can be
evaluated using the formulas~\eqref{eq:21}--\eqref{eq:22}:
\begin{multline}
  \label{eq:29}
  \left(\ldots \otimes \left\langle \varnothing, \left( t/q \right)^{3/4} w \right|\right)   (1 \otimes 1 \otimes \Delta) (1 \otimes \Delta) \mathcal{R}|_{\mathcal{F}^{(0,1)}(w) \otimes \mathcal{F}^{(1,0)}(u_1) \otimes
    \mathcal{F}^{(1,0)}(u_2) \otimes
    \mathcal{F}^{(1,0)}(u_3)}(|\square,w\rangle
\otimes \ldots) =\\
  =   \left(\ldots \otimes \left\langle \varnothing, \left( t/q \right)^{3/4} w \right|\right)
  (1 \otimes 1 \otimes \Delta) (1 \otimes \Delta) (P
  \mathcal{R}_0\mathcal{R}_1)\times\\
  \times (1
  \otimes 1 \otimes \Delta) (1 \otimes \Delta) \mathcal{R}_2  (|\square,w \rangle
\otimes \ldots).
\end{multline}

Using explicit formulas~\eqref{eq:60}--\eqref{eq:82} for the coproduct
and the Fock representations from Appendix~\ref{sec:representations}
we find that the second line of Eq.~\eqref{eq:29} is given by
\begin{equation}
  \label{eq:31}
  \left(\ldots \otimes \left\langle \varnothing, \left( t/q \right)^{3/4} w \right|\right)
  (1 \otimes 1 \otimes \Delta) (1 \otimes \Delta) (P
  \mathcal{R}_0\mathcal{R}_1) =  \mathcal{T}^{\varnothing}_{\varnothing}(w)  ( \langle
  \varnothing, w| \otimes \ldots ).
\end{equation}
Notice that the product $P \mathcal{R}_0 \mathcal{R}_1$ acts
diagonally on the state with empty Young diagram on the vertical line,
but exchanges the two tensor factors (due to $P$ operator) and shifts
the spectral parameter of the representation from $w (t/q)^{3/4}$ to
$w$ (due to the $\mathcal{R}_0$ piece involving grading
operators). What remains is to evaluate the third line of
Eq.~\eqref{eq:29}:
\begin{multline}
  \label{eq:33}
  ( \langle \varnothing, w| \otimes \ldots )\times (1 \otimes 1
  \otimes \Delta) (1 \otimes \Delta) \mathcal{R}_2 (|\square,w \rangle
  \otimes \ldots) =\\
  = \kappa_1 \oint \frac{dz}{z} \langle \varnothing, w|
  x^{-}(z)|\square,w \rangle \otimes (1 \otimes \Delta) \Delta
  \left(x^{+}(z)\right)=-\frac{\kappa_1}{1-t} (1 \otimes \Delta)
  \Delta \left(x^{+}(w)\right),
\end{multline}
where we have used the explicit action of $x^{-}(z)$ in the vertical
representation
\begin{equation}
  \label{eq:34}
  x^{-}(z)|\square,w \rangle = - \frac{1}{1-t}\delta \left( \frac{w}{z} \right) |\varnothing,w \rangle
\end{equation}
which follows from the definition~\eqref{eq:69}. In the second line of
Eq.~\eqref{eq:33} we have also used the fact that the identity term in
the definition~\eqref{eq:22} of $\mathcal{R}_2$ does not contribute to
the matrix element since it cannot turn $|\square\rangle$ into
$|\varnothing\rangle$. A more nontrivial observation is that all the
higher terms from Eq.~\eqref{eq:22} also don't contribute: they all
subtract two or more boxes from the diagram $|\square\rangle$ so that
it vanishes.

Collecting the terms from Eqs.~\eqref{eq:31} and \eqref{eq:33} we find
that
\begin{multline}
  \label{eq:83}
    \mathcal{R}|_{\mathcal{F}^{(0,1)}(w) \otimes
    \mathcal{F}^{(1,0)}(u_1)}
  \mathcal{R}|_{\mathcal{F}^{(0,1)}(\sqrt{t/q} w) \otimes
    \mathcal{F}^{(1,0)}(u_2)} \mathcal{R}|_{\mathcal{F}^{(0,1)}(t/q w)
    \otimes \mathcal{F}^{(1,0)}(u_3)} \sim \\
  \sim \left.  \mathcal{T}^{\varnothing}_{\varnothing}(w) (1 \otimes \Delta)
  \Delta \left(x^{+}(w)\right) \right|_{\mathcal{F}^{(1,0)}(u_1)\otimes
    \mathcal{F}^{(1,0)}(u_2)\otimes \mathcal{F}^{(1,0)}(u_3)}
\end{multline}

This explains why we get the coproduct of the $x^{+}(w)$ current from
the brane crossing~\eqref{eq:13} and ensures that the resulting
operator is indeed the $qq$-character, or $qW_m$-algebra generator as
shown in~\cite{MMZ} (and recalled in
Appendix~\ref{sec:qw_n-reduction}).

In the next section we will see that an insertion of the vertical
brane~\eqref{eq:13} into a toric diagram corresponding to a gauge
theory gives the Nekrasov formulas for the $qq$-characters.

\section{Normal ordering and commutation relations}
\label{sec:norm-order-comm}

To connect with the works of Nekrasov on the gauge theory origin of
$qq$-characters~\cite{Nekrasov:2017gzb}, we need to insert the
$qq$-character operator into a toric diagram corresponding to a gauge
theory. The resulting expressions will have the form of sums
(averages) over Young diagrams, or instanton series with brane
insertion producing an extra contribution to the measure.

Since the vertical brane insertion~\eqref{eq:13} is essentially a sum
of several free field vertex operators, its contribution to a refined
topological string partition function is easy to calculate using
Wick's theorem. The expressions for the refined topological vertices
in terms of free boson generators are written out in
Eqs.~\eqref{eq:35}, \eqref{eq:55}.

The normal ordering of $\mathcal{R}^{\lambda}_{\mu}(w)$
(Eq.~\eqref{eq:3}) with a product of $n$ $\Psi$-type and $m$
$\Psi^{*}$-type operators is given by:
\begin{enumerate}
\item $\lambda = \varnothing$, $\mu = \varnothing$:
\begin{equation}
  \label{eq:32}
\mathcal{R}^{\varnothing}_{\varnothing}(w)  \normord{ \prod_{a=1}^n \Psi^{\lambda^{(a)}}(x_a) \prod_{b=1}^m
  \Psi^{*}_{\mu^{(b)}}(y_b)} \,= \,\normord{ \mathcal{R}^{\varnothing}_{\varnothing}(w)  \prod_{a=1}^n \Psi^{\lambda^{(a)}}(x_a) \prod_{b=1}^m
  \Psi^{*}_{\mu^{(b)}}(y_b)},
\end{equation}
i.e.\ the normal ordering coefficient is trivial.

\item $\lambda = \square$, $\mu = \varnothing$:
  \begin{align}\nonumber
    \label{eq:37}
    &\mathcal{R}^{\square}_{\varnothing}(w) \normord{\prod_{a=1}^n
    \Psi^{\lambda^{(a)}}(x_a) \prod_{b=1}^m \Psi^{*}_{\mu^{(b)}}(y_b)}\,
     \,\\\nonumber
    &=\exp \Biggl[ \sum_{k \geq 1} \sum_{a=1}^n \frac{1}{k}\left(
      \frac{x_a}{w} \right)^k \left( 1 - (1-q^k)(1-t^{-k})
      \mathrm{Ch}_{\lambda^{(a)}}(q^k, t^k)
    \right)\\\nonumber&- \sum_{k \geq 1} \sum_{b=1}^m \frac{1}{k}\left(
      \sqrt{\frac{t}{q}} \frac{y_b}{w} \right)^k \left( 1 -
      (1-q^k)(1-t^{-k}) \mathrm{Ch}_{\mu^{(b)}}(q^k, t^k) \right)
    \Biggr] \normord{\mathcal{R}^{\square}_{\varnothing}(w) \prod_{a=1}^n
    \Psi^{\lambda^{(a)}}(x_a) \prod_{b=1}^m
    \Psi^{*}_{\mu^{(b)}}(y_b)} \\
    &= \frac{\mathsf{Y}_{\vec{\mu}}\left( \sqrt{\frac{q}{t}} w
      \right)}{\mathsf{Y}_{\vec{\lambda}}(w)}\,\normord{ \mathcal{R}^{\square}_{\varnothing}(w) \prod_{a=1}^n
    \Psi^{\lambda^{(a)}}(x_a) \prod_{b=1}^m
    \Psi^{*}_{\mu^{(b)}}(y_b)} 
  \end{align}
  with the so-called $\mathsf{Y}$-functions defined as
  \begin{multline}
    \label{eq:39}
    \mathsf{Y}_{\vec{\lambda}}(w) = \prod_{a=1}^n
    \mathsf{Y}_{\lambda^{(a)}}(w) \stackrel{\mathrm{def}}{=}
    \prod_{a=1}^n \left[\left( 1 - \frac{x_a}{w} \right) \prod_{(i,j)
        \in \lambda^{(a)}} \frac{\left( 1 - \frac{x_a}{w} q^{j}
          t^{1-i} \right) \left( 1 - \frac{x_a}{w} q^{j-1} t^{-i}
        \right)}{\left( 1 - \frac{x_a}{w} q^{j-1} t^{1-i} \right)
        \left( 1 -\frac{x_a}{w}
          q^{j} t^{-i}  \right)} \right]=\\
    =
    \prod_{a=1}^n\frac{N_{\lambda^{(a)}\square}\left(v^2\frac{x_{a}}{w}
      \right)}{N_{\lambda^{(a)}\varnothing}\left(v^2\frac{x_{a}}{w}
      \right)}
  \end{multline}
  where $n$ (the number of $\Psi$ or $\Psi^{*}$ intertwiners on a
  given horizontal brane) corresponds to the rank of the gauge group
  $U(n)$ on a given node of the quiver theory and $N_{\lambda \mu}(x)$
  is the Nekrasov factor given by
  Eq.~\eqref{eq:14}. $\mathsf{Y}$-functions~\eqref{eq:39} are natural
  observables in gauge theory~\cite{Nekrasov:2017gzb}.
  
\item $\lambda = \square$, $\mu = \square$:
  \begin{equation}
    \label{eq:38}
    \mathcal{R}^{\square}_{\square}\normord{\prod_{a=1}^n \Psi^{\lambda^{(a)}}(x_a) \prod_{b=1}^m
  \Psi^{*}_{\mu^{(b)}}(y_b)} \,= \,\normord{ \mathcal{R}^{\square}_{\square}(w)  \prod_{a=1}^n \Psi^{\lambda^{(a)}}(x_a) \prod_{b=1}^m
  \Psi^{*}_{\mu^{(b)}}(y_b)}.
  \end{equation}
  In this case again no normal ordering corrections are produced.
\end{enumerate}

In Eqs.~\eqref{eq:32}--\eqref{eq:39} the vertical brane insertion has
been placed to the left of all the other vertices. However, we can
also do a similar computation using Wick's theorem for the brane
inserted at any other point in the ``background'' brane diagram. It
would be more economical, however, to investigate directly the
commutation relations of the vertical brane insertion of the
formx~\eqref{eq:13} with other vertical edges on a toric diagram.

The crucial observation~\cite{Z20} is that due to the properties of
the $R$-matrix and the vertices, the vertical brane commutes with
internal edges:
\begin{equation}
  \label{eq:40}
  \includegraphics[valign=c]{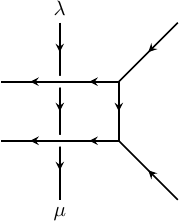} \quad =
  \quad \includegraphics[valign=c]{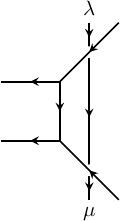}
\end{equation}
This can also be verified by a direct computation which we perform in
sec.~\ref{sec:refin-vert-comp}.

Another consequence of the properties of the $R$-matrix is that the
commutation of the vertical brane with external semi-infinite vertical
legs of the diagram produce only an inessential prefactor which does
not depend on the Young diagrams residing on the \textit{internal}
edges of the toric diagram.

In the next section we present the simplest example a $qq$-character
computed using the operator formalism we have developed.

\section{Fundamental $qq$-characters from the operator
  formalism and refined topological strings}
\label{sec:fund-qq-char}
In this section, we apply our operator formalism to explicitly
calculate the $qq$-characters in different theories. We also do the
same calculation in topological string theory using the formalism of
refined vertex~\cite{RefTopVert}.

\subsection{Fundamental $qq$-character of $A_1$ type}
\label{sec:fund-qq-char-1}

\subsubsection{Operator formalism}
\label{sec:operator-formalism}
The $A_1$ type $qq$-character can be obtained by inserting a vertical
spectator brane into the brane diagram corresponding to the toric
diagram of the local $\mathbb{F}_0$ or local $\mathbb{F}_1$ for gauge
theory with Chern-Simons level 0 and 1, respectively. The calculation
of the instanton partition function of the $U(2)$ theory using the
intertwining operators are briefly sketched in
Appendix~\ref{sec:algebr-engin-su2} for both cases.

\paragraph{$A_1$ $qq$-character in pure gauge theory without the
  Chern-Simons term.}
\label{sec:chern-simons-level}
Let us calculate the instanton partition function in the presence of
the vertical brane with no Chern-Simons term. We can use the normal
ordering identities~\eqref{eq:32},~\eqref{eq:37} and~\eqref{eq:38} to
obtain the partition function with the spectator brane insertion:
\begin{multline}
  \label{eq:41}
  \includegraphics[valign=c]{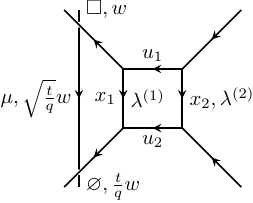}\quad \propto \quad
  \sum_{\vec{\lambda}} \left( \frac{t}{q} \frac{u_2 x_1}{u_1 x_2}
  \right)^{|\vec{\lambda}|}
  z_{\mathrm{vect}}(\vec{\lambda},\vec{x}) \Biggl(\,
  \underbrace{ \mathsf{Y}_{\vec{\lambda}} \left( \sqrt{\frac{q}{t}} w
    \right)}_{\mu= \varnothing} + \underbrace{ \left( \frac{x_1^2
        u_2}{w^2 u_1} \right) \frac{1}{\mathsf{Y}_{\vec{\lambda}}
      \left( \sqrt{\frac{t}{q}} w
      \right)}}_{\mu= \square} \,\Biggr)=\\
  = \sum_{\vec{\lambda}} \mathfrak{q}^{|\vec{\lambda}|}
  z_{\mathrm{vect}}(\vec{\lambda},\vec{x}) \Biggl(\,
  \underbrace{ \mathsf{Y}_{\vec{\lambda}} \left( \sqrt{\frac{q}{t}} w
    \right)}_{\mu= \varnothing} + \mathfrak{q} \underbrace{ \left(
      \frac{q x_1 x_2 }{t w^2 } \right)
    \frac{1}{\mathsf{Y}_{\vec{\lambda}} \left( \sqrt{\frac{t}{q}} w
      \right)}}_{\mu= \square} \,\Biggr)=\\
  = \left\langle \mathsf{Y} \left( \sqrt{\frac{q}{t}} w \right)
  \right\rangle + \mathfrak{q} \left( \frac{q x_1 x_2 }{t w^2 }
  \right) \left\langle \frac{1}{\mathsf{Y} \left( \sqrt{\frac{t}{q}} w
      \right)} \right\rangle = \chi_{qq}^{A_1}\left(\mathbb{C}^2\left|
      \sqrt{\frac{q}{t}}w \right.\right),
\end{multline}
where $\mathfrak{q} =\frac{t}{q} \frac{u_2 x_1}{ u_1 x_2}$ is the
instanton counting parameter and
\begin{equation}
  \label{eq:42}
  z_{\mathrm{vect}}(\vec{\lambda},\vec{x}) = \prod_{a,b=1}^2\frac{1}{N_{\lambda^{(a)}\lambda^{(b)}} \left( \frac{x_a}{x_b} \right)}
\end{equation}
is the vector multiplet contribution for $U(2)$ to the instanton
partition function. In the last line of Eq.~\eqref{eq:41} the average
is understood as the sum over pairs of Young diagrams with weight
$z_{\mathrm{vect}}(\vec{\lambda},\vec{x})$. Eq.~\eqref{eq:41} exactly
reproduces the fundamental $qq$-character of $A_1$ type~\cite{FR} (see
also~\cite{Nekrasov:2017gzb}, \cite{Kimura:2015rgi}).

\paragraph{$A_1$ $qq$-character in a theory with Chern-Simons level
  1.}
\label{sec:chern-simons-level-1}
Let us find the $qq$-character of the $5d$ gauge theory with the
Chern-Simons term of level 1 using the operator formalism. The
modification compared to Eq.~\eqref{eq:41} is minor: the contractions
between the $R$-matrix and the intertwining operators are the same,
only the prefactors change. We obtain the following $qq$-character for
Chern-Simons level 1:
\begin{equation}
  \includegraphics[raise=-1.5cm]{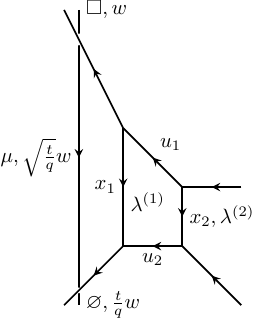} =  \left\langle \mathsf{Y}\left(\sqrt{\frac{q}{t}} w
    \right)\right\rangle + \mathfrak{q} \left(\frac{q x_1 x_2}{t
      w^3}\right) \left\langle
    \frac{1}{\mathsf{Y}\left(\sqrt{\frac{t}{q}} w \right)}\right\rangle
  = \chi_{qq}^{A_1,\, \mathrm{CS}=1} \left( \mathbb{C}^2\left|
      \sqrt{\frac{q}{t}} w \right. \right).
\label{cs1}
\end{equation}

\paragraph{$A_1$ $qq$-character in a theory with hypermultiplets.}
\label{sec:qq-character-theory-1}
The algebraic formalism allows us to compute the $qq$-character with
the matter hypermultiplets coupled to the gauge theory. As an example,
we will compute the fundamental character of $A_1$ in the presence of
four fundamental matter hypermultiplets. The contractions we have
already computed are enough to evaluate the character:
\begin{multline}
\includegraphics[valign=c]{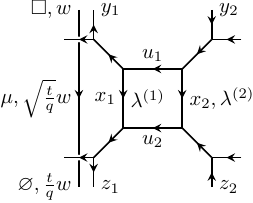}
\quad = \quad \left\langle \Ysf\left (\sqrt{\frac{q}{t}} w\right)
\right\rangle+\mathfrak{q}\left(\frac{qx_1x_2}{ty_1
    y_2}\right)P(w)\left\langle\frac{1}{\Ysf\left(
      \sqrt{\frac{t}{q}}w\right)}\right\rangle =\\
=\chi_{qq}^{A_1, \,
  4\mathrm{fund}}\left( \mathbb{C}^2\left| \sqrt{\frac{q}{t}} w
  \right. \right),\label{eq:93}
\end{multline}
where $P(w)$ is a new factor accounting for the matter multiplets that
depend on their masses encoded in the spectral parameters $z_{1,2}$,
$y_{1,2}$,
\begin{equation}
P(w) = \left(1-\frac{z_1}{w}\right)\left(1-\frac{z_2}{w}\right)\left(1-\frac{y_1}{w}\right)\left(1-\frac{y_2}{w}\right).\label{eq:94}
\end{equation}
The instanton counting parameter $\mathfrak{q}$ is also slightly
modified in the presence of matter fields,
\begin{equation}
  \label{eq:95}
  \mathfrak{q}=\frac{t u_2 x_1 y_2}{qu_1x_2z_1}.
\end{equation}

\subsubsection{Refined vertex computation}
\label{sec:refin-vert-comp}
In~\cite{AFS} the refined topological vertex has been shown to be
equal to the matrix element of certain intertwining operator of the
DIM algebra $\mathcal{A}$. This provided an algebraic approach to
compute the topological string amplitudes on local toric Calabi-Yau
threefolds. Using this relation, we want to show how the
$qq$-characters can be geometrically engineered. To introduce our
approach we will treat our first example, the $qq$-character of the
fundamental representation of $A_1$ using $U(2)$ theory in five
dimension, in some detail.

The $qq$-characters have been computed using the geometric transition
before~\cite{Kimura:2017auj}. In that work the authors focused on
geometries that initially engineer a higher rank gauge theory and
studied open topological string amplitudes by taking certain limits of
the appropriate K\"{a}hler classes corresponding to the geometric
transition, thus obtaining a lower rank theory with an operator
insertion. A different limit needed to be taken for each term in the
$qq$-character, in other words, the character was a linear combination
of amplitudes on different geometries. Although we study the open
topological amplitudes too, our approach is different in the sense
that all terms of the $qq$-character are produced from a single
geometry as one or more infinite sums over Young diagrams truncate to
a finite number of terms. To the best of our knowledge, this type of
truncation is used for the first time in the literature.

\begin{figure}[h]
  \begin{center}
    \includegraphics{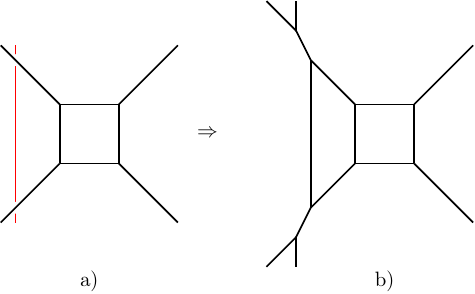}
  \end{center}
  \caption{a) The toric diagram of $\mathbb{F}_0$ corresponding to pure
    $U(2)$ gauge theory with vertical spectator brane corresponding to
    the $qq$-character insertion shown in red. b) The crossings of the
    red brane are resolved into conifold geometries, which gives rise
    to the toric diagram for $U(3)$ gauge theory with one fundamental
    and one anti-fundamental hypermultiplet. The red brane crossing
    from a) is recovered for certain degenerate values of K\"ahler
    parameters.}
  \label{transition}
\end{figure}

\begin{figure}[h]
  \begin{center}
    \includegraphics{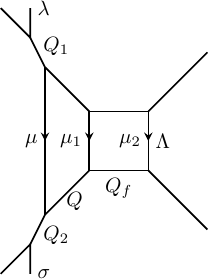}
  \end{center}
  \caption{Toric diagram corresponding to the $U(3)$ gauge theory with
    two extra hypermultiplets from Fig.~\ref{transition}~b) with
    complexified K\"ahler parameters written over the corresponding
    compact two-cycles. There is a summation over the Young diagrams
    $\mu$, $\mu_1$ and $\mu_2$ living on the internal edges, while the
    diagrams $\lambda$ and $\sigma$ on the external legs are
    parameters of the open topological string amplitude. The
    fundamental $A_1$ $qq$-character is obtained by setting $Q_1 =
    Q_2^{-1} = v^{-1}$ and $\lambda = \square$, $\sigma =
    \varnothing$.}
  \label{A1fund}
\end{figure}

\paragraph{$A_1$ $qq$-character in pure gauge theory without the
  Chern-Simons term.}
\label{sec:brane-inters-from}
For the $qq$-character of fundamental representation of $U(2)$ theory
with Chern-Simons level 0, we start with the toric threefold that
engineers $U(3)$ theory with two matter multiplets, one in the
fundamental and the other in the anti-fundamental
representation. Recalling the algebraic approach from
sec.~\ref{sec:operator-formalism}, this should be expected: inserting
an extra D5 brane would correspond to an increase in the rank of the
gauge theory and possibly new matter multiplets upon resolving
singularities. In addition, having a non-trivial states along the
additional brane translates into computing ``open'' amplitudes,
depicted in Figure~\ref{A1fund}. We obtain the $qq$-character by
imposing a ``degeneration'' condition in accordance with the
definition of the $R$-matrix Eq.~\eqref{eq:11}. We note that the
external legs of the toric diagram engineering the $U(3)$ theory with
two matter multiplets extend asymptotically in the same way as in the
geometry which engineers the $U(2)$ gauge theory with Chern-Simons
level 0 (local ${\mathbb F}_0$) depicted in
Fig.~\ref{transition}~a). Fig.~\ref{transition}~b) shows the toric
diagram after resolving the singularity.

The refined topological string partition function corresponding to
Fig.~\ref{A1fund} takes the following form after some slight
modifications to match it identically to instanton
counting,\footnote{We set the preferred direction to vertical in all
  the refined topological string computations throughout this paper.}
\begin{multline}
Z_{\lambda\sigma}\sim \sum_{\mu, \mu_1, \mu_2}\left(v^{-2}\Lambda Q_{f}^{-1} \right)^{|\mu_2|+|\mu_1|}\left( v^{-1}\Lambda Q^2\right)^{|\mu|}f_{\mu}^{-3}\left[ N_{\mu_1\mu_1}(1)N_{\mu_2\mu_2}(1)N_{\mu_2\mu_1}(Q_f)N_{\mu_1\mu_2}(Q_{f}^{-1})\right]^{-1}\times\\
\times \left[N_{\mu\mu}(1) N_{\mu_2\mu}(Q_f Q)N_{\mu_2\mu}(v^2Q_f
  Q)N_{\mu_1\mu}(Q) N_{\mu_1\mu}(v^2 Q)\right]^{-1}N_{\mu_2\lambda}(v
Q_f Q Q_1)N_{\mu_1\lambda}(v Q Q_1)\times \\
\times N_{\mu\lambda}(v Q_1)N_{\mu_2\sigma}(v Q_f Q Q_2)N_{\mu_1\sigma}(v Q Q_2)N_{\mu\sigma}(vQ_2),
\end{multline}
where $f_{\mu}$ is defined in Eq.~\eqref{eq:89}, $N_{\lambda \mu}(x)$ is
the Nekrasov factor Eq.~\eqref{eq:14} and
\begin{equation}
  \label{eq:88}
  v = \sqrt{\frac{q}{t}}.
\end{equation}
To obtain the $qq$-character we need to set the external
representations to $\lambda=\square$ and $\sigma=\varnothing$ and pick
special ``degenerate'' values for the K\"ahler parameters
\begin{equation}
  \label{eq:87}
  Q_1=v^{-1}, \qquad Q_2=v,
\end{equation}
which correspond to the geometric transtion from
Fig.~\ref{transition}. The sum over Young diagrams $\mu$ truncates to
a sum consisting of only two terms due to the identity
\begin{align}
N_{\mu\square}(1)=0,\qquad \text{unless } \mu\in\{\varnothing,\square\},
\end{align}
which we prove in Appendix~\ref{sec:appendix}. Note that the
$qq$-character of the fundamental representation of $A_1$ has indeed
two terms, as we also argued in the algebraic approach in
sec.~\ref{sec:operator-formalism}. In what follows we show that the
extra factors arising from the brane insertion give rise to the
$\Ysf$-operators. The remaining instanton sums are used to evaluate
their expectation values.

\begin{enumerate}[label=(\roman*)]
\item Consider $\mu = \varnothing$. After some cancellations we end up
  with the following additional factors under the gauge theory
  instanton sum:
\begin{equation}
  N_{\varnothing\square}(1)
  \frac{N_{\mu_1\square}(Q)}{N_{\mu_1\varnothing}(Q)}
  \frac{N_{\mu_2\square}(Q_{f}Q)}{N_{\mu_2\varnothing}(Q_{f}Q)} = \Ysf_{\mu_1}(vw)\Ysf_{\mu_2}(vw),
\end{equation} 
where $\mathsf{Y}_{\mu}(w)$ is defined in Eq.~\eqref{eq:39}.

We need to make an identification between the physical parameters of
the gauge theory and the geometric quantities, K\"{a}hler parameters,
used in the vertex computation. Using the following identification,
\begin{equation}
  Q_f=\frac{x_2}{x_1},\qquad \Lambda=\frac{u_2}{u_1}, \qquad
  Q=\frac{vx_1}{w},\label{eq:62}
\end{equation}
it can be seen that the first term in the $qq$-character that we had
found earlier is indeed reproduced (up to the
$N_{\varnothing\square}(1)$, which as we will show now gives only an
overall factor).

\item The second and last term in the sum over $\mu$ is when
  $\mu=\square$. This time we get
\begin{equation}
	(-1)v^{2}\left(v^{-1}\Lambda Q^2\right)f_{\square}^{-3} \; N_{ \varnothing\square}(1) \frac{N_{\mu_1\varnothing}(v^2 Q)}{N_{\mu_1\Box}(v^2Q)} \frac{N_{\mu_2\varnothing}(v^2Q_{f}Q)}{N_{\mu_2\Box }(v^2Q_{f}Q)} = \mathfrak{q}\left(\frac{q x_1 x_2}{tw^2} \right)\frac{1}{\Ysf_{\mu_1}(v^{-1}w)\Ysf_{\mu_2}(v^{-1}w)},
\end{equation}
which is identical to the second term in the $qq$-character we had
found earlier. We would like to emphasize that this result is the
``open'' topological string amplitude of a single toric diagram,
although we compute it in two steps: first the (two-term) sum over
$\mu$ and then over $\mu_1$, $\mu_2$.
\end{enumerate}

We can alternatively choose a different degeneration scheme by setting
$\lambda=\varnothing$ and $\sigma=\square$. The selection rule for
$\mu$ then follows from another choice of the K\"ahler classes $Q_1=v$
and $Q_2=v^{-1}$. It can be shown that in that case the same
$qq$-character is reproduced. One can argue more generally that the
second degeneration scheme gives the $qq$-characters for
\emph{conjugate} representations, so for the fundamental
representation of $A_1$ which is isomorphic to its conjugate the
answer remains the same.

\begin{figure}[h]
  \begin{center}
    \includegraphics{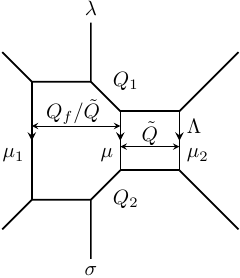}
  \end{center}
  \caption{Another toric diagram corresponding to $U(3)$ gauge theory
    with two extra hypermultiplets obtained from Fig.~\ref{A1fund} by
    commuting the vertical spectator brane with the internal edge carrying the
    Young diagram $\mu_1$. Notice the appearance of the new K\"ahler parameter $\tilde{Q}$.
  }
  \label{u2fund}
\end{figure}

\paragraph{$A_1$ $qq$-character in pure gauge theory without the
  Chern-Simons term. Alternative spectator brane position.}
\label{sec:comm-vert-brane}
In sec.~\ref{sec:norm-order-comm} we have noted that from general
algebraic arguments it follows that the vertical spectator brane
corresponding to the $R$-matrix commutes with internal edges of the
toric diagram. Now we would like to demonstrate it by a direct
computation in refined topological string theory. If we commute the
vertical brane with the internal edges we end up with a different
toric diagram depicted in Fig.~\ref{u2fund}. However, if we properly
modify the identification between the algebraic and geometric
variables after the commutation, we end up with the same
$qq$-character.

The open topological string amplitude corresponding to
Fig.~\ref{u2fund} is again expressed in terms of three infinite sums
over Young diagrams which corresponds to gluing two strip geometries
together. This amplitude can be manipulated to take the form of the
instanton counting of the $U(2)$ theory, with respect to which we will
compute the expectation values of the $\Ysf$-operators. Setting the
K\"{a}hler parameters $Q_1$, $Q_2$ to the degenerate
values~\eqref{eq:87} with the same choice of boundary conditions as
before, $\lambda=\square$ and $\sigma=\varnothing$, we get
\begin{multline}
  Z_{\square\varnothing}\sim \sum_{\vec{\mu}} \; (v^{-2}\Lambda Q_f^{-1})^{|\mu_2|+|\mu_1|} (v^{-1}\Lambda)^{|\mu|}\;f_\mu^{-1} \Big [N_{\mu_1\mu_1}(1)  N_{\mu_2\mu_2}(1)N_{\mu_2\mu_1}(Q_{f}) N_{\mu_1\mu_2}(Q_{f}^{-1}) \Big ]^{-1}\\
  \times \Big [ N_{\mu\mu}(1) N_{\mu_2\mu}(Q) N_{\mu_2\mu}(v^2 Q)
  N_{\mu\mu_1}(Q_{f}Q^{-1}) N_{\mu\mu_1}(v^2Q_{f}Q^{-1}) \Big]^{-1}
  N_{\mu_2\square}(Q)N_{\mu\square}(1)N_{\square\mu_1}(v^2Q_{f}Q^{-1})
  \\
  \times N_{\mu_2\varnothing}(v^2Q)N_{\mu_2\varnothing}(v^2)
  N_{\varnothing \mu_1}(Q_{f}Q^{-1}),
\end{multline}
which agrees with the $qq$-character~\eqref{eq:41} found earlier
provided that we identify
\begin{align}
\label{eqCS0}
  \tilde{Q} = \frac{vx_2}{w} = Q Q_f.
\end{align}

Let us note that we also get the same $qq$-character if we choose
$\lambda=\varnothing$ and $\sigma=\square$ with $Q_1= v$,
$Q_2=v^{-1}$.

\begin{figure}[h]
\begin{center}
\includegraphics{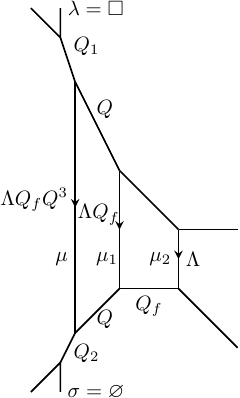}
\end{center}
\caption{Resolution of crossings in Eq.~\eqref{cs1} describing the
  vertical spectator brane insertion into the toric diagram of local
  $\mathbb{F}_1$ corresponding to $U(2)$ gauge theory with
  Chern-Simons level 1.}
\label{fig:A1CS1}
\end{figure}

\paragraph{$A_1$ $qq$-character in a theory with Chern-Simons level 1.}
\label{sec:a_1-qq-character}
We can compute the $qq$-character in the presence of a Chern-Simons
term using the topological vertex as well. The toric diagram is
depicted in Fig.~\ref{fig:A1CS1}. We start with the local
$\mathbb{F}_1$ which engineers the $U(2)$ theory at Chern-Simons
level 1 and insert a D5 brane as before.

The topological sting amplitude represented by Fig.~\ref{fig:A1CS1}
after taking $\lambda=\square$, $\sigma=\varnothing$, $Q_1=v^{-1}$ and
$Q_2=v$ becomes:
\begin{multline}
  Z\sim \sum_{\vec{\mu}}(v^{-2}\Lambda
  Q_f^{-1})^{|\mu_2|}(v^{-2}\Lambda)^{|\mu_1|}(v^{-1}\Lambda Q_f
  Q^3)^{|\mu|}f_{\mu_1}^{-1}f_{\mu_2}^{-1}f_{\mu}^{-4}\times
  \\
  \times \Big[N_{\mu_1\mu_1}(1)N_{\mu_2\mu_2}(1)N_{\mu_2\mu_1}(Q_f)N_{\mu_1\mu_2}(Q_f^{-1})\Big]^{-1} \Big[N_{\mu\mu}(1)N_{\mu_2\mu}(Q_f Q)N_{\mu_2\mu}(v^2Q_f Q) N_{\mu_1\mu}(Q)N_{\mu_1\mu}(v^2 Q)\Big]^{-1}\times\\
  \times N_{\mu_2\square}(Q_f
  Q)N_{\mu_1\square}(Q)N_{\mu\square}(1)N_{\mu_2\varnothing}(v^2 Q_f
  Q) N_{\mu_1\varnothing}(v^2Q)N_{\mu\varnothing}(v^2).\label{eq:90}
\end{multline}
We have almost the same map between the algebraic and geometric
parameters as in the absence of the Chern-Simons term
(cf.~Eq.~\eqref{eq:62}):
\begin{equation}
  Q_f=\frac{x_2}{x_1},\qquad \Lambda=-\frac{u_2}{u_1}x_2^{-1}, \qquad
  Q=\frac{vx_1}{w}, \label{eq:73}
\end{equation}
with the same instanton parameter $\mathfrak{q} =\frac{t}{q} \frac{u_2
  x_1}{ u_1 x_2}$. With the identification Eq.~\eqref{eq:73} the
refined topological string amplitude~\eqref{eq:90} reproduces the
character~\eqref{cs1}.

\begin{figure}[h]
	\begin{center}
          \includegraphics{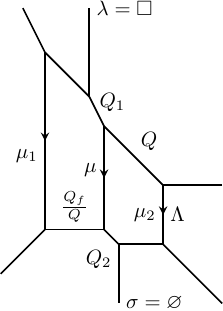}
        	\end{center}
                \caption{The toric diagram corresponding to an
                  alternative point of insertion of the vertical
                  spectator brane into the $\mathbb{F}_1$
                  geometry. The new K\"ahler parameter $\tilde{Q}$ is
                  related to $Q$ by Eq.~\eqref{eq:92}.}
\label{A1fundCS1}
\end{figure}

\paragraph{$A_1$ $qq$-character in a theory with Chern-Simons level
  1. Alternative spectator brane position.}
\label{sec:diff-brane-insert}
We argued in sec.~\ref{sec:norm-order-comm} that the exact point at
which the vertical brane (representing the $R$-matrix) is inserted
into a toric diagram does not affect the form of the $qq$-character,
but slightly modifies the identification between the K\"ahler
parameters of the CY and the spectral parameters of the DIM
representations. Let us demonstrate that the same is true for the case
of nontrivial Chern-Simons level as well. The toric diagram
corresponding to an alternative vertical brane insertion is depicted
in Fig.~\ref{A1fundCS1}.

The topological string amplitude is reduced to,
\begin{multline}
Z_{\square\varnothing}\sim \sum_{\vec{\mu}}(v^{-2}\Lambda Q_{f}^{-1})^{|\mu_2|}(v^{-2}\Lambda)^{|\mu_1|}(\Lambda Q)^{|\mu|}f_{\mu_1}^{-1}f_{\mu_2}^{-1}f_{\mu}^{-1}\Big[N_{\mu_1\mu_1}(1)N_{\mu_2\mu_2}(1)N_{\mu_2\mu_1}(Q_f)N_{\mu_1\mu_2}(Q_f^{-1})\Big]^{-1}\times\\
\times \Big [N_{\mu\mu}(1)N_{\mu_2\mu}(Q)N_{\mu_2\mu}(v^2Q)N_{\mu\mu_1}(Q_f Q^{-1}) N_{\mu\mu_1}(v^2Q_f Q^{-1}) \Big]^{-1}N_{\mu_2\square}(Q)N_{\mu\square}(1)N_{\square\mu_1}(v^2Q_f Q^{-1})\times\\
\times N_{\mu_2\varnothing}(v^2Q)N_{\varnothing\mu}(1)N_{\varnothing\mu_1}(Q_f Q^{-1}),\label{eq:91}
\end{multline}
when the boundary conditions are set to $\lambda=\square$ and
$\sigma=\varnothing$ with $Q_1=Q_2=v^{-1}$. With the alternative
insertion of the vertical brane we obtain the same $qq$-character as
in Eq.~\eqref{eqCS0} if we identify
\begin{equation}
  \tilde{Q}=\frac{vx_2}{w} = Q Q_f \label{eq:92}
\end{equation}
as in the case of vanishing Chern-Simons level.

The ``opposite'' degeneration scheme with $\lambda=\varnothing$,
$\sigma=\square$ is also possible for the non-vanishing Chern-Simons
level, and the two alternatives give rise to the same $qq$-character.

\begin{figure}[h!]
  \begin{center}
    \includegraphics{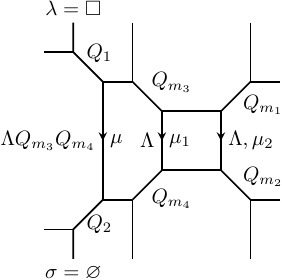}
  \end{center}
  \caption{Resolution of the crossings of vertical brane inserted into
    the toric diagram corresponding to the $U(2)$ gauge theory with
    four fundamental hypermultiplets.}
\label{A14fundvert}
\end{figure}

\paragraph{$A_1$ $qq$-character in a theory with hypermultiplets.}
\label{sec:qq-character-theory}
As in the previous cases in this section, the $qq$-character with
additional matter hypermultiplets can be obtained from refined
topological strings. The corresponding toric diagram is shown in
Fig.~\ref{A14fundvert}.

We set the boundary conditions as before, $\lambda=\square$ and
$\sigma=\varnothing$ with $Q_1=v^{-1}$ and $Q_2=v$, so that the
topological string partition functions reduces to
\begin{multline}
Z_{\square\varnothing}\sim \sum_{\vec{\mu}}\left(v^{-2}\Lambda Q_{m_1}Q_{m_4}\right)^{|\mu_1|+|\mu_2|}\left(v^{-1}\Lambda Q_{m_3}Q_{m_4}\right)^{|\mu|}f_{\mu}^{-1}\Big[N_{\mu_1\mu_1}(1)N_{\mu_2\mu_2}(1)N_{\mu_2\mu_1}(Q_f)N_{\mu_1\mu_2}(Q_f^{-1})\Big]^{-1}\times\\
\times N_{\mu_2\varnothing}(v Q_{m_1}^{-1})N_{\mu_1\varnothing}(v
Q_{f}^{-1}Q_{m_1}^{-1})N_{\mu_2\varnothing}(v Q_f
Q_{m_3})N_{\mu_1\varnothing}(v Q_{m_3})N_{\varnothing\mu_2}(v
Q_{m_2})N_{\varnothing\mu_1}(vQ_f Q_{m_2})\times \\
\times N_{\varnothing\mu_2}(vQ_f^{-1}
Q_{m_4}^{-1})N_{\varnothing\mu_1}(vQ_{m_4}^{-1})\Big[
N_{\mu\mu}(1)N_{\mu_1\mu}(Q)N_{\mu_1\mu}(v^2 Q)N_{\mu_2\mu}(Q_f
Q)N_{\mu_2\mu}(v^2 Q_f Q)\Big]^{-1} \times \\
\times N_{\varnothing\mu}(vQ_f Q Q_{m_1})N_{\varnothing\mu}(vQ_f Q
Q_{m_2})N_{\varnothing\mu}(vQ Q_{m_3}^{-1})N_{\varnothing\mu}(vQ
Q_{m_4}^{-1})N_{\mu_2\square}(Q_f
Q)N_{\mu_1\square}(Q)N_{\mu\square}(1) \times\\
\times N_{\mu_2\varnothing}(v^2 Q_f Q)N_{\mu_1\varnothing}(v^2 Q)N_{\mu\varnothing}(v^2).
\end{multline}

We obtain a match with the $qq$-character~\eqref{eq:93} by setting the
map between the algebraic and geometric parameters as follows
\begin{equation}
Q_{f}=\frac{x_2}{x_1},\qquad \Lambda=\frac{u_2}{u_1},\qquad Q_{m_1}=\frac{y_2}{x_2},\qquad Q_{m_2}=\frac{z_2}{x_2},\qquad Q_{m_3}=\frac{x_1}{y_1},\qquad Q_{m_4}=\frac{x_1}{z_1}.
\end{equation}

\subsection{Fundamental $qq$-character of $A_2$ type}
\label{sec:fund-qq-char-2}

\subsubsection{Operator formalism}
\label{sec:operator-formalism-1}

\begin{figure}[h]
  \begin{center}
    \includegraphics{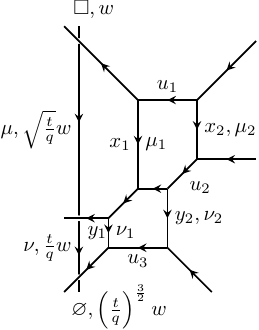}
  \end{center}
  \caption{The network of intertwiners reproducing to the fundamental
    $A_2$ $qq$-character.}
  \label{A2DIMfund}
\end{figure}

Our formalism also produces $qq$-characters for higher rank $A$-type
root systems. The procedure is similar to the $A_1$ case described in
sec.~\ref{sec:fund-qq-char-1}. We start with a network of DIM
intertwiners corresponding to a quiver gauge theory of type $A_n$,
insert the $R$-matrix into it and take its matrix element. We would
like to demonstrate this approach on a simple example of $A_2$. The
network of intertwiners is depicted in Fig.~\ref{A2DIMfund}.

Leaving out the details of the calculation, we give the answer for the
matrix element of the network from Fig.~\ref{A2DIMfund} which indeed
has the form of the $A_2$ $qq$-character:
\begin{equation}
  \chi_{qq}^{A_2} \left( \mathbb{C}^3 \left| \sqrt{\frac{q}{t}}
      w\right. \right)=\left
    \langle\Ysf_1\left(\sqrt{\frac{q}{t}}w\right)\right\rangle+\left(\frac{u_2
      x_1^2}{u_1 y_1 y_2
      w}\right)\left\langle\frac{\Ysf_2(w)}{\Ysf_1\left(
        \sqrt{\frac{t}{q}}w\right)}\right\rangle+\left(\sqrt{\frac{q}{t}}\frac{u_3x_1
      y_1}{u_1 w^2} \right)\left\langle
    \frac{1}{\Ysf_2\left(\frac{t}{q}w \right)}\right\rangle.
\end{equation}

\subsubsection{Refined vertex computation}
\label{sec:refin-vert-comp-1}

\begin{figure}[h]
  \begin{center}
     \includegraphics{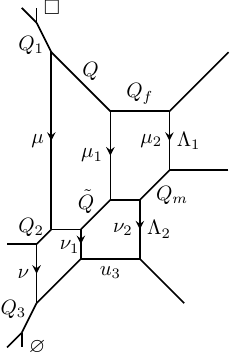}
   \end{center}
   \caption{The toric diagram obtained from Fig.~\ref{A2DIMfund} by
     resolving all the crossings. This gives rise to a quiver
     $U(3)\times U(3)$ gauge theory.}
   \label{A2vertfund}
\end{figure}

The same $A_2$ $qq$-character as in
sec.~\ref{sec:operator-formalism-1} can also be calculated following
the refined topological vertex approach. The toric diagram with the
$R$-matrix insertion is drawn in Fig.~\ref{A2vertfund}.

The degeneration needed to reproduce the $qq$-character occurs at
$Q_1=v^{-1}$ and $Q_2=Q_3=v$. The map between the K\"ahler parameters
and the spectral parameters is given by
\begin{equation}
Q_f=\frac{x_2}{x_1},\qquad Q_{m}=\frac{x_2}{y_2},\qquad\widetilde{Q}=\frac{x_1}{y_1},\qquad\Lambda_1=-\frac{u_2}{u_1}x_2^{-1},\qquad\Lambda_2=\frac{u_3}{u_2}y_2,\qquad Q=\frac{v x_1}{w}.\label{eq:96}
\end{equation}

\section{Multiple vertical brane insertions and higher
  $qq$-characters}
\label{sec:higher-qq-characters}
In this section we consider several vertical ``spectator'' branes
(representing the $R$-matrices) inserted into the same network of
intertwiners. We show that this gives rise to $qq$-characters of
higher representations of $A_m$ algebras. As in
sec.~\ref{sec:fund-qq-char} we first derive our results in the
algebraic approach employing DIM intertwining operators and then give
the parallel ``geometric'' computation using refined topological
vertex formalism.

\subsection{Operator formalism}
\label{sec:operator-formalism-2}
To understand how to combine several vertical branes together let
us start with the simplest example with two vertical ``spectator''
branes intersecting two horizontal lines:
\begin{equation}
  \label{eq:43}
    \includegraphics[valign=c]{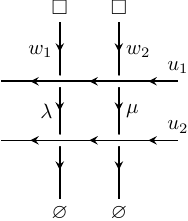}\quad \sim
    \quad \left. \mathcal{T}^{\varnothing}_{\varnothing}(w_1)
    \Delta(x^{+}(w_1)) \mathcal{T}^{\varnothing}_{\varnothing}(w_2) \Delta(x^{+}(w_2))\right|_{\mathcal{F}^{(1,0)}_{q, t^{-1}}(u_1) \otimes
    \mathcal{F}^{(1,0)}_{q, t^{-1}}(u_2)}
\end{equation}
where $\mathcal{T}^{\varnothing}_{\varnothing}(w)$ has been defined in
Eq.~\eqref{eq:24} and $\Delta(x^{+}(w))$ is given by
Eq.~\eqref{eq:60}. Using the formulas Eqs.~\eqref{eq:63}--\eqref{eq:77}
for the horizontal Fock representation and the normal ordering
identities from sec.~\ref{sec:norm-order-comm} we find that
\begin{gather}
  \label{eq:45}
  \mathcal{T}^{\varnothing}_{\varnothing}(w_1)
  \mathcal{T}^{\varnothing}_{\varnothing}(w_2) =\, :\mathcal{T}^{\varnothing}_{\varnothing}(w_1)
  \mathcal{T}^{\varnothing}_{\varnothing}(w_2):\\
  \left.\Delta(x^{+}(w_1))\right|_{\mathcal{F}^{(1,0)}_{q, t^{-1}}(u_1) \otimes
    \mathcal{F}^{(1,0)}_{q, t^{-1}}(u_2)}
  \mathcal{T}^{\varnothing}_{\varnothing}(w_2) = \frac{1 - \frac{t}{q}
  \frac{w_2}{w_1}}{1-\frac{w_2}{w_1}} :\left.\Delta(x^{+}(w_1))\right|_{\mathcal{F}^{(1,0)}_{q, t^{-1}}(u_1) \otimes
    \mathcal{F}^{(1,0)}_{q, t^{-1}}(u_2)}
  \mathcal{T}^{\varnothing}_{\varnothing}(w_2):, \\
  \mathcal{T}^{\varnothing}_{\varnothing}(w_1) \left.\Delta(x^{+}(w_2))\right|_{\mathcal{F}^{(1,0)}_{q, t^{-1}}(u_1) \otimes
    \mathcal{F}^{(1,0)}_{q, t^{-1}}(u_2)} =\, :\mathcal{T}^{\varnothing}_{\varnothing}(w_1) \left.\Delta(x^{+}(w_2))\right|_{\mathcal{F}^{(1,0)}_{q, t^{-1}}(u_1) \otimes
    \mathcal{F}^{(1,0)}_{q, t^{-1}}(u_2)}:.
\end{gather}

\begin{multline}
  \label{eq:46}
  \left.\Delta(x^{+}(w_1))\right|_{\mathcal{F}^{(1,0)}_{q,
      t^{-1}}(u_1) \otimes \mathcal{F}^{(1,0)}_{q, t^{-1}}(u_2)}
  \left.\Delta(x^{+}(w_2))\right|_{\mathcal{F}^{(1,0)}_{q,
      t^{-1}}(u_1) \otimes
    \mathcal{F}^{(1,0)}_{q, t^{-1}}(u_2)} \sim\\
  \sim\, :x^{+}(w_1) x^{+}(w_2): \otimes 1 + \mathcal{S}\left(
    \frac{w_1}{w_2} \right) :x^{+}(w_1) \psi^{-} \left( \left(
      \frac{t}{q} \right)^{\frac{1}{4}}w_2
  \right): \otimes\, x^{+} \left( \sqrt{\frac{t}{q}}w_2 \right) +\\
  + S \left( \frac{w_2}{w_1} \right) :\psi^{-} \left( \left(
      \frac{t}{q} \right)^{\frac{1}{4}}w_1
  \right) x^{+}(w_2) : \otimes\, x^{+}\left(\sqrt{\frac{t}{q}} w_1
  \right) +\\
  + :\psi^{-} \left( \left(
      \frac{t}{q} \right)^{\frac{1}{4}}w_1
  \right) \psi^{-} \left( \left(
      \frac{t}{q} \right)^{\frac{1}{4}}w_2
  \right)  : \otimes : x^{+}\left( \sqrt{\frac{t}{q}} w_1 \right)
  x^{+}\left( \sqrt{\frac{t}{q}} w_2 \right) :,
\end{multline}
where the ``structure function'' $S(x)$ is defined as
\begin{equation}
  \label{eq:47}
  S(x) \stackrel{\mathrm{def}}{=} \frac{\left( 1 - \frac{x}{t} \right) ( 1 - q x
    )}{(1-x) \left( 1 - \frac{q}{t} x \right)}.
\end{equation}
Inserting the resulting operator Eq.~\eqref{eq:43} into a toric diagram
corresponding to a pure $SU(2)$ gauge theory gives
\begin{multline}
  \label{eq:48}
 \includegraphics[valign=c]{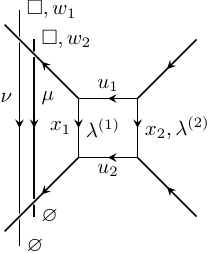}\quad \sim \quad 
  \underbrace{\left\langle   \mathsf{Y}_{\vec{\lambda}} \left( \sqrt{\frac{q}{t}}
        w_1 \right) \mathsf{Y}_{\vec{\lambda}} \left(
        \sqrt{\frac{q}{t}} w_2
      \right) \right\rangle}_{\nu= \varnothing,\, \mu= \varnothing} 
    + \underbrace{\mathfrak{q}\, S \left( \frac{w_2}{w_1} \right)
     \left( \frac{q x_1 x_2 }{t w^2_1 } \right)
      \left\langle \frac{\mathsf{Y}_{\vec{\lambda}} \left( \sqrt{\frac{q}{t}} w_2
        \right)}{\mathsf{Y}_{\vec{\lambda}} \left( \sqrt{\frac{t}{q}}
          w_1 \right)}\right\rangle}_{\nu= \square,\, \mu=
    \varnothing}+\\
  +
    \underbrace{\mathfrak{q}\, S \left( \frac{w_1}{w_2} \right)
     \left( \frac{q x_1 x_2 }{t w^2_2 } \right)
      \left\langle \frac{\mathsf{Y}_{\vec{\lambda}} \left( \sqrt{\frac{q}{t}} w_1
        \right)}{\mathsf{Y}_{\vec{\lambda}} \left( \sqrt{\frac{t}{q}}
          w_2 \right)}\right\rangle}_{\nu= \varnothing,\, \mu=
    \square} + \underbrace{\mathfrak{q}^2  \left( \frac{q^2 x_1^2 x_2^2 }{t^2
          w^2_1 w^2_2 } \right) \left\langle \frac{1}{\mathsf{Y}_{\vec{\lambda}}
        \left( \sqrt{\frac{t}{q}} w_1 \right)
        \mathsf{Y}_{\vec{\lambda}} \left( \sqrt{\frac{t}{q}} w_2
        \right)}\right\rangle}_{\nu= \square,\, \mu= \square} 
  =\\
  =\chi_{qq}^{A_1}\left(S^2\mathbb{C}^2\left|
      \sqrt{\frac{q}{t}}w_1, \sqrt{\frac{q}{t}}w_2 \right.\right),
\end{multline}
which exactly reproduces weight $2$ $qq$-character of $A_1$ type
corresponding to symmetric representation~\cite{Kimura:2015rgi}.

To get the $qq$-character corresponding to antisymmetric
representation $\Lambda^2 \mathbb{C}^2$ of $A_1$ (which is trivial) we
need to consider a degeneration of Eq.~\eqref{eq:48}:
\begin{equation}
  \label{eq:49}
  \chi_{qq}^{A_1}\left(\Lambda^2\mathbb{C}^2\left|
      \sqrt{\frac{q}{t}}w_1 \right.\right) \sim
  \lim_{w_2 \to \frac{t}{q} w_1} \chi_{qq}^{A_1}\left(S^2\mathbb{C}^2\left|
      \sqrt{\frac{q}{t}}w_1, \sqrt{\frac{q}{t}}w_2 \right.\right).
\end{equation}
The value $\frac{w_2}{w_1} = \frac{t}{q}$ is a pole of the function
$S(\frac{w_2}{w_1})$, therefore, to get a meaningful answer
we have to multiply the character under the limit by something
proportional to $\left( 1 - \frac{q}{t} \frac{w_2}{w_1}\right)$. Only
one term survives:
\begin{equation}
  \label{eq:50}
  \lim_{w_2 \to \frac{t}{q} w_1} \left( 1 - \frac{q}{t}
    \frac{w_2}{w_1}\right) \chi_{qq}^{A_1}\left(S^2\mathbb{C}^2\left|
      \sqrt{\frac{q}{t}}w_1, \sqrt{\frac{q}{t}}w_2 \right.\right) \sim  \underbrace{
      \left\langle \frac{\cancel{\mathsf{Y}_{\vec{\lambda}} \left(
            \sqrt{\frac{q}{t}} \frac{t}{q} w_1
        \right)}}{\cancel{\mathsf{Y}_{\vec{\lambda}} \left( \sqrt{\frac{t}{q}}
          w_1 \right)}}\right\rangle}_{\nu= \square,\, \mu=
    \varnothing} = 1,
\end{equation}
which is indeed equal to the $qq$-character of the trivial
representation $\Lambda^2 \mathbb{C}^2$ of $A_1$.

The degeneration pattern of higher $qq$-characters Eq.~\eqref{eq:50} is in
fact general. For example, one can get the $qq$-characters of all the
fundamental representations $\Lambda^k \mathbb{C}^n$ of $A_{n-1}$ by
considering $k$ vertical branes with coordinates $w_i$, $i=1,\ldots,k$
intersecting $n$ horizontal lines in the toric diagram and sending
\begin{equation}
  \label{eq:51}
  w_i = \left( \frac{t}{q} \right)^{i-1} w.
\end{equation}

It is curious to notice which Young diagrams on the intermediate legs
(the analogs of $\nu$ and $\mu$ in Eq.~\eqref{eq:48}) survive the
degeneration. The pattern of intermediate diagrams is shown below
\begin{equation}
  \label{eq:52}
  \chi_{qq}^{A_n} \left( \Lambda^k \mathbb{C}^n| w \right) \sim \includegraphics[valign=c]{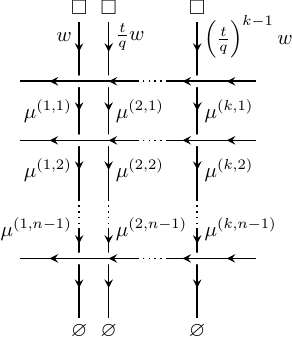}.
\end{equation}
The only sets of $\mu^{(i,j)} = \square$ or $\varnothing$ which give
nontrival contributions to the $qq$-character Eq.~\eqref{eq:52} are
\begin{gather}
  \label{eq:53}
  \mu^{(i+1,j)} \subseteq \mu^{(i,j)}, \qquad \mu^{(i,j+1)} \subseteq
  \mu^{(i,j)},\\
  \sum_{i=1}^k \mu^{(i,j)} - \sum_{i=1}^k \mu^{(i,j+1)} = 0 \text{ or
  } 1,\label{eq:97}
\end{gather}
where we set $\mu^{(i,0)} \stackrel{\mathrm{def}}{=} \square$ and $\mu^{(i,n)} \stackrel{\mathrm{def}}{=}
\varnothing$. Let us give an example of an allowed pattern of
$\mu^{(i,j)}$'s for $n=6$, $k=4$
\begin{equation}
  \label{eq:54}
  \begin{array}{cccc}
    \square & \square &\square &\square \\
    \square & \square &\square &\square \\
    \square & \square &\square &\varnothing \\
    \square & \square  &\varnothing &\varnothing \\
    \square &\varnothing  &\varnothing &\varnothing \\
    \square &\varnothing  &\varnothing &\varnothing    
  \end{array}
\end{equation}
It is evident that for $k > n$ there are no allowed configurations of
$\mu^{(i,j)}$. Indeed, at the top of the pattern there is a row of $k$
boxes, while at the bottom there is an empty row, and at every step
down at most one box is eliminated. There is no way the total of $k$
boxes can be eliminated after $n$ steps. This fits with the fact that
the corresponding representations $\Lambda^k \mathbb{C}^n$ vanish
identically.

After a bit of experimentation one can deduce that the number of
allowed configurations Eqs.~\eqref{eq:53},~\eqref{eq:97} is given by
$\frac{n!}{k! (n-k)!}$, which is precisely the dimension of $\Lambda^k
\mathbb{C}^n$.

\subsection{Refined vertex computation}
\label{sec:refin-vert-comp-2}

\begin{figure}[h]
	\begin{center}
		\includegraphics{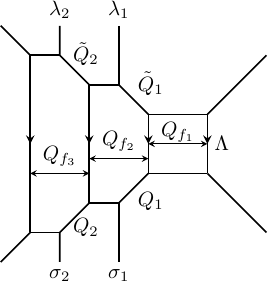}
        \end{center}
        \caption{Toric diagram geometrically engineering a $U(4)$
          gauge theory with two fundamental and two antifundamental
          hypermultiplets. After the degeneration of the K\"ahler
          parameters $Q_1$, $Q_2$, $\tilde{Q}_1$, $\tilde{Q}_2$ the
          open string amplitude reproduces the $qq$-character of the
          symmetric representation $S^2\mathbb{C}^2$ of $A_1$.}
\label{u2sym}
\end{figure}

The geometric approach involving refined topological vertices allows
us to calculate the $qq$-characters of higher representations too. As
an illustrative example, we calculate the character of the symmetric
representation of $A_1$. We insert branes at two different locations
associated with two parameters of the $qq$-character. The
corresponding toric Calabi-Yau threefold before the degeneration of
the K\"{a}hler parameters, whose toric diagram is depicted in Figure
\ref{u2sym}, engineers a $U(4)$ gauge theory with two fundamental and
two anti-fundamental matter hypermultiplets.

To get the $qq$-character we calculate the open topological string
amplitude for $\lambda_1=\lambda_2=\varnothing$ and
$\sigma_1=\sigma_2=\square$ with the K\"{a}hler parameters set to
degenerate values $Q_1=Q_2=v^{-1}$ and
$\widetilde{Q}_1=\widetilde{Q}_2=v$.  Due to the degeneration two out
of four infinite sums over all Young diagrams in the instanton sum of
the $U(4)$ theory truncate so that each of them consists of only two
terms making up the four terms in the character:
\begin{multline}
  Z^{\varnothing\varnothing}_{\Box\Box} \sim v^2\, z_1^2
  z_2^2 \Big\langle\Ysf(v^2 z_1)\Ysf(v^2 z_2)\Big\rangle
  +\textgoth{q}\,v\,
  z_2^2\,S\left(\frac{z_2}{z_1}\right)\Big\langle\frac{\Ysf(v^2
    z_2)}{\Ysf(z_1)}\Big\rangle+\textgoth{q}\,v\,
  z_1^2\,S\left(\frac{z_1}{z_2}\right)\Big\langle\frac{\Ysf(v^2
    x_1)}{\Ysf(z_2)}\Big\rangle +\\
  +\mathfrak{q}^2\Big\langle\frac{1}{\Ysf(z_1)\Ysf(z_2)}\Big\rangle,\label{eq:98}
\end{multline}
where the ``structure function'' $S(x)$ is the same as in
Eq.~\eqref{eq:47}. In the calculation leading to Eq.~\eqref{eq:98} the
following alternative formula for $S(x)$ through the
Nekrasov factors Eq.~\eqref{eq:14} is useful:
\begin{equation}
  S(x)=\frac{N_{\Box\Box}(x)}{N_{\Box\varnothing}(x)N_{\varnothing\Box}(x)}. \label{eq:99}
\end{equation}
The instanton counting parameter in Eq.~\eqref{eq:98} is given by
\begin{equation}
  \label{eq:100}
  \mathfrak{q}=v^2 \Lambda Q_f^{-1},
\end{equation}
while the map between the K\"ahler parameters and the parameters of
hte representation is similar to~\eqref{eq:62}.

We should compare our expression~\eqref{eq:98} with Eq.~(4.18)
of~\cite{Bourgine:2017jsi}.  At first sight, they may look different,
however, they turn out to be identical after combining the terms due
to the normal ordering of the DIM operators with the rest of the
expression.

\section{Elliptic generalization of $qq$-characters}
\label{sec:elliptic-lift-qq}

Our algebraic approach can be generalized to produce the $6d$ version
of the $qq$-characters introduced
in~\cite{MMZ},~\cite{Kimura-ell}. This is achieved by taking the
elliptic deformation of the DIM algebra and intertwining
operators~\cite{saito}. We review the basics of the procedure in
Appendix~\ref{sec:elliptic-deformation}. The deformation is equivalent
to the ``partial compactification'' of the toric CY giving rise to an
elliptic fibration~\cite{Nieri-IKY}. M-theory compactified on the
fibration engineers the $6d$ gauge theory on $\mathbb{R}^4 \times
T^2$, hence the notion of $6d$ $qq$-characters.

\subsection{Operator formalism}
\label{sec:operator-formalism-3}

The intertwining operators are vertex operators involving free bosonic
modes $a_n$ obeying the $(q,t)$-deformed Heisenberg
algebra~\eqref{eq:86}. In the elliptic deformation of the DIM algebra
we introduce an extra parameter $p$ into the commutation
relations~\cite{saito},~\cite{Ghoneim:2020sqi}. It turns out that to
survive the deformation, the horizontal Fock representation defined in
Appendix~\ref{sec:representations} needs to be ``doubled'', i.e.\ one
needs to introduce the \emph{second set} of independent Heisenberg
generators $b_n$ in addition to $a_n$. More details are provided in
Appendix~\ref{sec:ellipt-deform-horiz}.

Due to the doubling of the modes the elliptic deformation of each
vertex operator becomes a product of two vertex operators: one
expressed in terms of the old Heisenberg generators $a_n$ and the
other in terms of the new ones $b_n$. The elliptic version of the
intertwining operator~\eqref{eq:35} takes the form\footnote{We denote
  the elliptic intertwining operators by $\Phi$ and $\Phi^{*}$ to
  distinguish them from the undeformed ones, which we call~$\Psi$,
  $\Psi^{*}$.},
\begin{multline}
\Phi^{\lambda}(x)=\quad \includegraphics[valign=c]{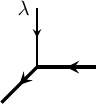} \quad\sim \, \normord{\exp\left[\sum_{n\neq 0}\frac{1}{n}\frac{1}{1-p^{|n|}}\left( \frac{1}{1-q^{-n}}-(1-t^n)\mbox{Ch}_{\lambda}(q^{-n},t^{-n})\right)x^{-n}a_n\right]}\times\\
\times \normord{\exp\left[ \sum_{n\neq 0} \frac{1}{n}\frac{p^{|n|}}{1-p^{|n|}}\left(\frac{1}{1-q^n}-(1-t^{-n})\mbox{Ch}_{\lambda}(q^n,t^n) \right)x^n b_n\right]},\label{eq:101}
\end{multline}
while its dual Eq.~\eqref{eq:55} becomes
\begin{multline}
\Phi^{*}_{\mu}(y)= \quad \includegraphics[valign=c]{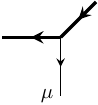} \quad\sim\, \normord{\exp\left[-\sum_{n\neq
      0}\frac{1}{n}\frac{q^{-|n|/2}t^{|n|/2}}{1-p^{|n|}}\left(
      \frac{1}{1-q^{-n}}-(1-t^n)\mbox{Ch}_{\mu}(q^{-n},t^{-n})\right)y^{-n}a_n\right]}
\times\\
\times \normord{\exp\left[ -\sum_{n\neq 0} \frac{1}{n}\frac{p^{|n|}}{1-p^{|n|}}q^{|n|/2}t^{-|n|/2}\left(\frac{1}{1-q^n}-(1-t^{-n})\mbox{Ch}_{\mu}(q^n,t^n) \right)y^n b_n\right]},\label{eq:102}
\end{multline}
where we denote the ``doubled'' Fock representations by thick lines,
$\mathrm{Ch}_{\lambda}(q,t)$ is defined by Eq.~\eqref{eq:36} and we
have skipped the prefactors which are the same as in the undeformed
case (see Eqs.~\eqref{eq:84}, \eqref{eq:85}).  The Heisenberg
generators $a_n$ and $b_n$ satisfy the elliptically deformed
commutation relations~\eqref{eq:105}. One can check that the vertex
operators~\eqref{eq:101}, \eqref{eq:102} are indeed the intertwining
operators between the vertical and horizontal Fock representations of
the elliptic DIM algebra as defined in
Appendix~\ref{sec:elliptic-deformation}.

We can use the deformed intertwining operators~\eqref{eq:101},
\eqref{eq:102} to compute the elliptic deformation of the
$R$-matrix~\eqref{eq:3} which we denote by $\mathfrak{R}$,
\begin{multline}
  \mathfrak{R}^{\lambda}_{\mu}(w,u,N) \stackrel{\mathrm{def}}{=}
  \Phi^{\lambda}(w) \Phi_{\mu}\left( \sqrt{\frac{t}{q}}w
  \right)=\quad \includegraphics[valign=c]{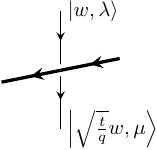}
  \quad  =\\
  = (-w)^{-N|\lambda|}u^{|\lambda|} \left( - \sqrt{\frac{t}{q}} w
  \right)^{N|\mu|} \left( \frac{q}{u} \sqrt{\frac{t}{q}}
  \right)^{|\mu|} \left( \frac{f_{\mu}}{f_{\lambda}} \right)^N
  f_\lambda^{-1} q^{n(\lambda^{\mathrm{T}})+
    n(\mu^{\mathrm{T}})}
  \prod_{k=1}^{\infty} N_{\lambda \mu}\left(p^{k-1}
    \frac{q}{t} \right)N_{\mu\lambda }\left(p^{k} \right) \times\\
  \times :\exp \Biggl[ - \sum_{n \geq 1} \frac{1}{n} \frac{1}{1-p^n}
  \frac{ 1- \left( t/q \right)^n}{1-q^n}w^n a_{-n} \Biggl]\times
  \\
  \times \exp\Biggl[ - \sum_{n \neq 0} \frac{1 }{n} \frac{1-t^n
  }{1-p^{|n|}}
  \Biggl( \mathrm{Ch}_{\lambda}(q^{-n},t^{-n}) - \left( \frac{t}{q}
  \right)^{\frac{|n|-n}{2}} \mathrm{Ch}_{\mu}(q^{-n},t^{-n})
  \Biggl)w^{-n} a_n \Biggl] \,: \times
  \\
  \times :\exp \Biggl[ \sum_{n \geq 1} \frac{1}{n} \frac{p^n}{1-p^n}
  \frac{q^n\left( 1- \left( q/t \right)^n\right)}{1-q^n} w^{-n}b_{-n}
  \Biggl] \times
  \\
  \times \exp\Biggl[ - \sum_{n \neq 0} \frac{1 }{n}
  \frac{p^{|n|}(1-t^{-n}) }{1-p^{|n|}}
  \Biggl( \mathrm{Ch}_{\lambda}(q^{n},t^{n}) - \left( \frac{q}{t}
  \right)^{\frac{|n|-n}{2}} \mathrm{Ch}_{\mu}(q^{n},t^{n})
  \Biggl)w^{n} b_n \Biggl] \,:,\label{eq:104}
\end{multline}
where the Nekrasov factors $N_{\lambda\mu}(x)$ are given by
Eq.~\eqref{eq:14}, $\mathrm{Ch}_{\lambda}$, $n(\lambda^{\mathrm{T}})$
and $f_{\lambda}$ are defined by Eqs.~\eqref{eq:36}, \eqref{eq:113}
and \eqref{eq:89} respectively.

The selection rule for $\lambda$ and $\mu$ in Eq.~\eqref{eq:104} is
the same as in sec.~\ref{sec:dim-r-matrices}: when $\lambda=\square$
the other representation $\mu$ can be either $\varnothing$ or
$\square$ for $\mathfrak{R}^{\lambda}_{\mu}$ to be nonzero. The
prefactor in the first line of Eq.~\eqref{eq:104} is slightly
different from that of the original $R$-matrix~\eqref{eq:3} --- it is
missing the $c_\lambda\,c_\mu$ factors. The reason behind this
difference is our choice of basis in the vertical Fock
representation. In the elliptic calculations in this section we use a
more convenient (orthogonal) basis in which the norm is given by
\begin{equation}
  \langle\lambda | \mu\rangle =\delta_{\mu\nu}(-1)^{|\lambda|}q^{\|\lambda\|^2/2+|\lambda|/2}t^{\|\lambda^{\mathrm{T}}\|^2/2-|\lambda|/2}\prod_{k=1}^{\infty} N_{\lambda\lambda}(p^{k-1})N_{\lambda\lambda}(p^k v^2).\label{eq:116}
\end{equation}
where $\|\lambda\|^2$ is defined by Eq.~\eqref{eq:115}.

The norm~\eqref{eq:116} is proportional to the vector multiplet
contribution to the Nekrasov partition function of the $6d$ $U(1)$
gauge theory. Moreover, for $p \to 0$ it reduces to the familiar
integral form normalization of Macdonald polynomials $J_{\lambda}$,
\begin{equation}
  \langle\lambda | \mu\rangle \stackrel{p\to 0}{\to} \langle J_\lambda
  | J_\mu\rangle =\delta_{\mu\lambda} c_\mu c'_\mu,\qquad
  \text{with}\qquad c'_{\mu}=\prod_{(i,j)\in\mu}\left(
    1-q^{\mu_i-j+1}t^{\mu^{\mathrm{T}}_j-i}\right).
\end{equation}

One can evaluate the Wick contractions of the $R$-matrix
Eq.~\eqref{eq:104} with the intertwining operators~\eqref{eq:101},
\eqref{eq:102} and show that the only non-trivial contributions (apart
from the possible prefactors) happens when $\lambda=\square$ and
$\mu=\varnothing$.  The contraction has the same form as in the
undeformed case (Eq.~\eqref{eq:37}) but with modified $\mathsf{Y}$
functions which we denote by $\mathcal{Y}$,
\begin{multline}
  \mathcal{Y}_{\vec{\lambda}}(w)= \prod_{a=1}^n
  \mathcal{Y}_{\lambda^{(a)}}(w) \equiv  \prod_{a=1}^n \prod_{k=1}^{\infty}\frac{N_{\lambda^{(a)}\square}\left(v^2p^{k-1}\frac{x_a}{w}\right)N_{\square\lambda^{(a)}}\left(p^{k-1}\frac{w}{x_a}\right)}{N_{\lambda^{(a)}\varnothing}\left(v^2p^{k-1}\frac{x_a}{w}\right)N_{\varnothing\lambda^{(a)}}\left(p^{k-1}\frac{w}{x_a}\right)}= \\
  =\prod_{a=1}^n \theta_p \left(\frac{x_a}{w}\right)
  \prod_{(i,j)\in\lambda^{(a)}} \frac{\theta_p
    \left(\frac{x_a}{w}q^{j-1}t^{-i}\right) \theta_p
    \left(\frac{x_a}{w}q^{j}t^{-i+1}\right)}{\theta_p
    \left(\frac{x_a}{w}q^{j}t^{-i}\right)
    \theta_p \left(\frac{x_a}{w}q^{j-1}t^{-i+1}\right)}.\label{eq:117}
\end{multline}
In the definition~\eqref{eq:117} we have used the Jacobi theta
function $\theta_p(x)$ (Eq.~\eqref{eq:112}) instead of the more
conventional one $\theta_1(x;p)$. The explicit expressions for the
$qq$-characters are the same up to a multiplicative factor in front of
them when written in terms of either of these functions.

\begin{figure}[h]
\begin{center}
  \includegraphics{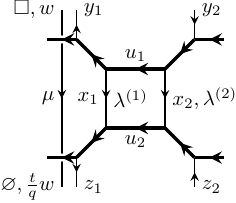}
  \caption{The vertical brane inserted into the network of
    intertwiners corresponding to the $6d$ $U(2)$ gauge theory with
    four hypermultiplets in fundamental representation. We denote the
    ``doubled'' Fock representations by thick lines.  }
\end{center}
\label{6dint}
\end{figure}

\paragraph{Fundamental elliptic $A_1$ $qq$-character.}
\label{sec:a_1-qq-character-1}
We would like to compute the elliptic (or $6d$) $qq$-character of the
fundamental representation for the $A_1$ group. In the $6d$ case to
avoid anomaly we need to make sure that the theory is conformal. The
simplest nonabelian gauge group is just $U(2)$, so to have vanishing
beta function we need to add four matter hypermultiplets in
fundamental representations. First, let us present the computation
using the elliptic deformation of the intertwining operators given by
Eqs.~\eqref{eq:101}, \eqref{eq:102}. The network of intertwiners
corresponding to the $6d$ theory is depicted in
Fig.~\ref{6dint}. Notice that the diagrams for the networks
corresponding to the $5d$ and $6d$ theories are the same, only the
intertwining operators $\Psi$, $\Psi^{*}$ are replaced with their
elliptic deformations $\Phi$, $\Phi^{*}$.

Similarly to the trigonometric case, we can normal order all the
operators (the $R$-matrices and triple intertwiners) in the network
and obtain the following $qq$-character:
\begin{equation}
  \chi_{qq}^{A_1,\, \mathrm{ell}} \left( \mathbb{C}^2\left|
      \sqrt{\frac{q}{t}} w \right. \right) \sim \left\langle \mathcal{Y}\left(\sqrt{\frac{q}{t}}w \right)\right\rangle + \mathfrak{q}\, \mathcal{P}(w)\left( \frac{q x_1 x_2 }{t z_1 y_1 }
  \right)\left\langle \frac{1}{\mathcal{Y}\left(\sqrt{\frac{t}{q}}w \right)}\right\rangle. \label{eq:118}
\end{equation}
Here, $\mathcal{P}(w)$ is the elliptic version of the factor $P(w)$
introduced in Eq.~\eqref{eq:94},
\begin{equation}
  \mathcal{P}(w)=\prod_{a=1}^{N_f}\theta_p\left(\frac{y_a}{w}\right),\label{eq:119}
\end{equation}
where $y$ is the insertion point of the intertwining operators
corresponding to the matter hypermultiplets. If the elliptic
deformation parameter $p$ is sent to zero, we obviously get back to
the trigonometric result from
sec.~\ref{sec:qq-character-theory-1}~\cite{Nekrasov:2017gzb},~\cite{MMZ}.

\begin{figure}[h]
  \begin{center}
    \includegraphics{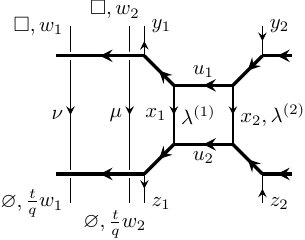}
  \end{center}
  \caption{Two vertical branes inserted into the network of
    intertwiners corresponding to the $6d$ $U(2)$ gauge theory with
    four hypermultiplets in fundamental representation. Only $\mu =
    \square, \varnothing$ and $\nu = \square, \varnothing$ produce
    nontrivial contributions leading to four terms in the
    $qq$-character~\eqref{eq:103}.}
  \label{fig:higher-ell}
\end{figure}

\paragraph{Higher elliptic $A_1$ $qq$-character.}
\label{sec:higher-a_1-qq}
The elliptic version of our algebraic approach can be extended to
multiple brane insertions as well. As an example we would like to
insert two $R$-matrices into the diagram corresponding to the $6d$
$U(2)$ gauge theory. The DIM network is depicted in
Fig.~\ref{fig:higher-ell}.

The network produces the elliptic $qq$-character corresponding to the
first symmetric representation of $A_1$:
\begin{multline}
  \chi_{qq}^{A_1, \, \mathrm{ell}} \left(S^2\mathbb{C}^2\left|
      \sqrt{\frac{q}{t}}w_1, \sqrt{\frac{q}{t}}w_2 \right.\right) =\left\langle \mathcal{Y}\left(\sqrt{\frac{q}{t}}w_1 \right) \mathcal{Y}\left(\sqrt{\frac{q}{t}}w_2 \right)\right\rangle+\left( \frac{u_2x_1^2}{u_1 y_1 z_1}\right)\mathcal{P}(w_2)\mathcal{S}\left(\frac{w_1}{w_2}\right)\left\langle\frac{ \mathcal{Y}\left(\sqrt{\frac{q}{t}}w_1 \right) }{\mathcal{Y}\left(\sqrt{\frac{t}{q}}w_2 \right)}\right\rangle+\\
  +\left( \frac{u_2x_1^2}{u_1 y_1
      z_1}\right)\mathcal{P}(w_1)\mathcal{S}\left(\frac{w_2}{w_1}\right)\left\langle\frac{
      \mathcal{Y}\left(\sqrt{\frac{q}{t}}w_2 \right)
    }{\mathcal{Y}\left(\sqrt{\frac{t}{q}}w_1 \right)}\right\rangle+\left(
    \frac{u_2x_1^2}{u_1 y_1 z_1}\right)^2
  \mathcal{P}(w_1)\mathcal{P}(w_2)\left\langle\frac{1
    }{\mathcal{Y}\left(\sqrt{\frac{t}{q}}w_1 \right)
      \mathcal{Y}\left(\sqrt{\frac{t}{q}}w_2 \right)}\right\rangle,
\label{eq:103}
\end{multline}
where $\mathcal{P}(w)$ is given by Eq.~\eqref{eq:119} and we have
defined the elliptic version $\mathcal{S}(w)$ of the structure
function $S(w)$ (Eq.~\eqref{eq:47}),
\begin{equation}
  \mathcal{S}(w)=\frac{\theta_p(qw)\theta_p(t^{-1}w)}{\theta_p(w)\theta_p(q t^{-1}w)}.
\end{equation}
Note that we have defined the elliptic versions of all the functions
featuring in the $qq$-characters using the Jacobi theta function
$\theta_p(x)$ instead of $\theta_1(x;p)$. It is easy to check that the
functional form of the $qq$-character would not have changed if we had
defined them using $\theta_1$.

\subsection{Refined vertex computation}
\label{sec:refin-vert-comp-3}

Let us also comment on an alternative derivation of the elliptic
$qq$-character~\eqref{eq:118}. As shown
in~\cite{Nieri-IKY},~\cite{Ghoneim:2020sqi}, the elliptic deformation
of a vacuum matrix element of a network of intertwiners can be
achieved by taking the trace of the network instead. To this end we
need to evaluate the matrix elements associated with external
horizontal lines (Fock representations) in the diagram from
Fig.~\ref{A14fundvert} of the form $\langle P_{\mu}|\ldots
|P_{\mu}\rangle$, where ``$\ldots$'' stand for a product of the
intertwining operators $\Psi$ and $\Psi^{*}$. We then perform the sum
over all possible Young diagrams $\mu$ with the weight $p^{|\mu|}$. To
get back the undeformed expression~\eqref{eq:93} one needs to take $p
\to 0$ so that only the vacuum matrix element $\langle
\varnothing|\mathellipsis |\varnothing\rangle$ contributes to the sum.

This approach is parallel to the refined topological vertex
computation in which one studies the ``compactified'' toric
diagram~\cite{Hollowood:2003cv}, i.e.\ the diagram drawn on a cylinder
instead of a plane. To be able to glue the external lines in the toric
diagram together one needs to ensure that their slopes match. This
turns out to be equivalent to requiring that the corresponding gauge
theory is superconformal (cf.\ the diagram which cannot be
compactified, e.g.\ from Fig.~\ref{fig:A1CS1}). Let us demonstrate how
it works in several examples.

\begin{figure}[h]
\begin{center}
\includegraphics{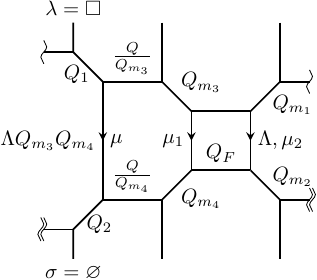}
\end{center}
\caption{The compactified toric diagram corresponding to $6d$ $U(3)$
  gauge theory with six fundamental hypermultiplets to make it
  superconformal. The wavy lines indicate that the corresponding
  points need to be identified. The partition function reproduces the
  elliptic $qq$-character in $6d$ $U(2)$ gauge theory provided one
  sets $Q_2 = Q_1^{-1} = v$.}
\label{fig:6dfundtoric}
\end{figure}

\paragraph{Fundamental elliptic $A_1$ $qq$-character and $6d$ $U(2)$
  gauge theory.}
\label{sec:fund-ellipt-a_1}
We have pointed out before that the $R$-matrix commutes with the
vertex operators, which manifests itself in different possible
degeneration schemes of the toric diagrams leading to the same
$qq$-character. In the following we would like to present two
different toric diagrams that give rise to the fundamental elliptic
$A_1$ $qq$-character. The first version is depicted in
Fig.~\ref{fig:6dfundtoric}.

After applying the degeneration condition $Q_1 = v^{-1}$, $Q_2 = v$
and fixing $\lambda=\square$, $\sigma=\varnothing$ one can show that
the refined topological string partition function reproduces the
elliptic $qq$-character~\eqref{eq:118} obtained in the algebraic
approach. In the process one also fixes the identification between the
K\"ahler parameters of the CY background and the spectral parameters
of Fock representations of the algebra $\mathcal{A}$:
\begin{equation}
  \label{dic}
  Q_f=\frac{x_2}{x_1},\qquad \Lambda=\frac{u_2}{u_1},\qquad
  Q_{m_1}=\frac{y_2}{x_2},\qquad Q_{m_2}=\frac{z_2}{x_2},\qquad
  Q_{m_3}=\frac{x_1}{y_1},\qquad Q_{m_4}=\frac{x_1}{z_1},
\end{equation}
and for the insertion point of the spectator brane one gets
\begin{equation}
Q=v\frac{x_1}{w}.\label{eq:120}
\end{equation}

\begin{figure}[h]
\begin{center}
\includegraphics{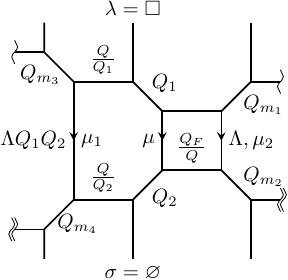}
\end{center}
\caption{Alternative degeneration of the toric diagram of the $6d$
  $U(3)$ gauge theory with six hypermultiplets,
  cf. Fig.~\ref{fig:6dfundtoric}. Degenerate values of the K\"ahler
  parameters reproducing the $qq$-character are $Q_2 = Q_1^{-1} = v$.}
\label{fig:6dtoric2}
\end{figure}

\paragraph{Fundamental elliptic $A_1$ $qq$-character and $6d$ $U(2)$
  gauge theory. Alternative spectator brane position.}
\label{sec:vert-brane-insert}
The second type of degeneration of the $U(3)$ toric diagram leading to
the elliptic $A_1$ $qq$-character is when the vertical spectator brane
is placed in the compact four cycle, as depicted in
Fig.~\ref{fig:6dtoric2}.
The topological string partition function is identical to the elliptic
$qq$-character with the same dictionary between spectral parameters
and K\"ahler parameters as in Eq.~\eqref{dic} and the slightly
different K\"{a}hler class used to parametrize the insertion of the
extra brane,
\begin{equation}
  Q=v^{-1}\frac{w}{x_1}.
  \label{eq:121}
\end{equation}

\paragraph{Higher elliptic $A_1$ $qq$-character and $6d$ $U(2)$ gauge
  theory.}
\label{sec:higher-elliptic-a_1}

\begin{figure}[h]
\begin{center}
 \includegraphics{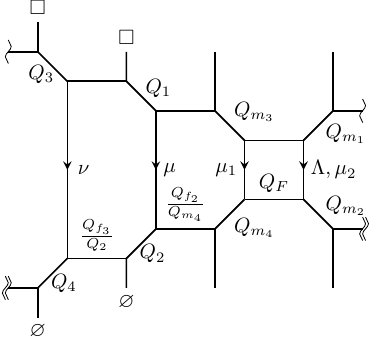}
\end{center}
\caption{Toric diagram corresponding to the $6d$ $U(4)$ gauge theory
  with eight fundamental hypermultiplets. The wavy lines indicate that
  the corresponding points need to be identified. The degeneration
  which gives rise to $qq$-character happens when one sets $Q_1 = Q_3
  = v^{-1}$, $Q_2 = Q_4 = v$.}
\label{fig:6dvertmulti}
\end{figure}

Again, we can reproduce the $qq$-character~\eqref{eq:103} obtained in
the algebraic approach from refined topological string. To this end we
should compute the amplitude corresponding to the diagram depicted in
Fig.~\ref{fig:6dvertmulti}.

Formula~\eqref{eq:103} is indeed reproduced provided we use the
dictionary~\eqref{dic} between the parameters and identify the
spectator brane insertions points with
\begin{align}
Q_{f_2}=\frac{vx_1}{w_2},\qquad\text{and}\qquad Q_{f_3}=\frac{w_2}{w_1}.
\end{align}

\section{Conclusions}
\label{sec:conclusions}
We have presented a new description of $qq$-characters using the
$R$-matrix of the DIM algebra. Our approach is inspired by the
correspondence between Type IIB branes and representations of the DIM
algebra~\cite{AFS},~\cite{Z20} in which the $R$-matrix is viewed as a
spectator brane insertion. This allows for a compact uniform
description of all $qq$-characters of the $A_n$ series both for
fundamental (defining) representation and higher representations.  We
also provide an uplift of the formalism to elliptic $qq$-characters.

Several further lines of investigation are apparent. In the present
paper we have considered the insertions of the spectator brane with
external states $|\square\rangle$ and $|\varnothing\rangle$ only. One
can ask if other choices of external states also correspond to some
$qq$-characters. The preliminary answer is affirmative, in particular
we believe that the \emph{exchange} of the two external states leads
to the $qq$-character of the conjugate representation.

A more radical generalization would be to consider the spectator brane
that is not a 5-brane, but a 7-brane. It is reasonable to conjecture
that in this case the MacMahon $R$-matrix put forward
in~\cite{Awata:2018svb} should appear, giving rise to new
$qq$-characters. There is a large class of DIM representations arising
from degenerations of the MacMahon representation which includes the
among others the Fock representation. The MacMahon $R$-matrix will
allow one to study $qq$-characters corresponding to this class of DIM
representations.

The uniform treatment of the $A_n$ series of $qq$-characters resembles
the uniform approach of~\cite{Bazhanov:2005as} to quantum affine Lie
algebras. In that case the integrals of motion for all
$U_q(\widehat{A}_n)$ algebras are obtained uniformly in $n$ by
building the transfer-matrices from a more elementary
higher-dimensional object, the Zamolodchikov tetrahedron operator. It
would be interesting to investigate if the $qq$-characters can be
viewed in the same way with certain tetrahedron operator replacing the
DIM $R$-matrix.

From a gauge theory perspective it would be desirable to study the
connection of our current approach with the work~\cite{Haouzi:2019jzk}
where the $qq$-characters were interpreted as partition functions of a
coupled $1d$-$5d$ field theory system. The $1d$ degrees of freedom
represent the Wilson line in a $5d$ gauge theory which looks similar
to how the spectator brane producing the $qq$-character is coupled to
the 5-brane system in our $R$-matrix formalism. This connection
definitely merits further investigation.

Recently the algebraic approach to $qq$-characters using DIM algebra
has been extended to gauge theories with gauge groups from the
classical series $B_n$, $C_n$ and $D_n$~\cite{Nawata:2023wnk}. We
believe that our $R$-matrix formalism can also be extended to these
cases putting them in the unified framework with the $A_n$ examples
presented here. We plan to study such an extension in our next works.



%
%
%
%

\section*{Acknowledgments}
YZ is grateful to CK and to Bo\u{g}azi\c{c}i University where this
work has been initiated. CK would like to thank the warm hospitality
of Niels Bohr Institute where this work was finished. This work is
partly supported by the joint grant RFBR 21-51-46010 and
T\"{U}B\.{I}TAK 220N106. CK's research is also supported by
T\"{U}B\.{I}TAK grants 120F184. Mehmet Batu Bayındırlı is supported by
T\"{U}B\.{I}TAK grant 220N106. This article is based upon work from
COST Action 21109 CaLISTA, supported by COST (European Cooperation in
Science and Technology).

\appendix

\section{Basic facts about the DIM algebra}
\label{sec:basic-facts-about}
In this Appendix we give a short summary of the Ding-Iohara-Miki (DIM)
algebra $U_{q,t}(\widehat{\widehat{\mathfrak{gl}}}_1)$. We will skip
many technical details which can be found e.g.\ in~ \cite{MMZ},
\cite{Z18}, \cite{Z21}.

\paragraph{Definition.}
\label{sec:definition}
The algebra $U_{q,t}(\widehat{\widehat{\mathfrak{gl}}}_1)$ can be
viewed as the quantum deformation of the universal enveloping algebra
of a Lie algebra $qW_{1+\infty}$ generated by the elements $e_{(n,m)}$
  with $(n,m) \in \mathbb{Z}^2$ satisfying
\begin{equation}
  \label{eq:57}
    [e_{(n,m)}, e_{(k,l)}] = \left( q^{\frac{nl-mk}{2}} -
      q^{-\frac{nl-mk}{2}} \right) e_{(n+k,m+l)} + (c_1 n + c_2 m)
      \delta_{n+k,0} \delta_{m+l,0},
\end{equation}
where $c_{1,2}$ are two central elements. The Lie algebra
$qW_{1+\infty}$ is doubly graded with $e_{(n,m)}$ having grading
$(n,m)$. We denote the grading operators by $d_1$ and $d_2$
respectively:
\begin{equation}
  \label{eq:19}
  [d_1,e_{(n,m)}] = n e_{(n,m)}, \qquad   [d_2,e_{(n,m)}] = m e_{(n,m)}.
\end{equation}
The automorphism group $SL(2,\mathbb{Z})$ acts on
$qW_{1+\infty}$ so that the indices $(n,m)$ of the generators
$e_{(n,m)}$ and the central charges $(c_1, c_2)$ transform in the
fundamental two-dimensional representation and its conjugate
respectively.

The quantum deformation $U_{q,t}(\widehat{\widehat{\mathfrak{gl}}}_1)$
of $qW_{1+\infty}$ is controlled by the parameter $t/q$. When $t/q$ is
equal to one, the Lie algebra relations~\eqref{eq:57} are
reproduced. We will keep the notation $e_{(n,m)}$ for the elements of
the quantum deformed universal enveloping algebra corresponding to the
elements of the Lie algebra, though the relations between them will
become nonlinear after the deformation. We will not write these
complicated relations explicitly here (see e.g.~\cite{Z21}).

The double grading survives the quantum deformation, while the
$SL(2,\mathbb{Z})$ action is uplifted to the action of its universal
cover $\widetilde{SL}(2,\mathbb{Z}) = SL(2,\mathbb{Z}) \ltimes
\mathbb{Z}$. Another nontrivial property of the algebra
$U_{q,t}(\widehat{\widehat{\mathfrak{gl}}}_1)$ is that the parameter
$q$ of $qW_{1+\infty}$, the deformation parameter $t/q$ and their
inverse product $t^{-1}$ enter the commutation relations only as
symmetric combinations.

The $\widetilde{SL}(2,\mathbb{Z})$ automorphism group
and the symmetry under the permutation of the deformation parameters
$(q,t^{-1}, t/q)$ imply that the set of representations of
$U_{q,t}(\widehat{\widehat{\mathfrak{gl}}}_1)$ is organized into
orbits of the $\widetilde{SL}(2,\mathbb{Z}) \times \mathfrak{S}_3$
group action.

\paragraph{Coproduct(s).}
\label{sec:coproducts}
The DIM algebra is a nontrivial Hopf algebra. To write out the
coproduct it will be convenient to introduce the generating currents
\begin{align}
  \label{eq:58}
  x^{\pm}(z) &= \sum_{n \in \mathbb{Z}} e_{(\pm 1,n)}z^{-n},\\
  \psi^{\pm}(z) & = K^{\mp 1} \exp \left(- \sum_{n \geq
      1}\frac{\kappa_n}{n} \gamma^{\mp \frac{n}{2}} e_{(0,\pm n)} z^{\mp
      n} \right),\label{eq:59}
\end{align}
where $\kappa_n = (1-q^n)(1-t^{-1}) (1 - (t/q)^n)$, and $\gamma =
(t/q)^{c_1/2}$ and $K = (t/q)^{c_2/2}$ are the central element of the
DIM algebra corresponding to the central elements $c_1$ and $c_2$ of
$qW_{1+\infty}$ respectively.

In terms of the currents~\eqref{eq:58}, \eqref{eq:59} the coproduct
$\Delta$ can be written as
\begin{align}
  \label{eq:60}
    \Delta(x^{+}(z)) &= x^{+}(z) \otimes 1 + \psi^{-}\left(
    \gamma_{(1)}^{\frac{1}{2}}z \right) \otimes x^{+} \left(
    \gamma_{(1)} z \right),\\
  \Delta(x^{-}(z)) &= x^{-}\left(\gamma_{(2)} z\right) \otimes
  \psi^{+}\left( \gamma_{(2)}^{\frac{1}{2}}z \right) + 1 \otimes
  x^{-}(z),\\
  \Delta(\psi^{\pm}(z)) &= \psi^{\pm}\left(
    \gamma_{(2)}^{\pm \frac{1}{2}}z \right) \otimes \psi^{\pm}\left(
    \gamma_{(1)}^{\mp \frac{1}{2}}z \right),\label{eq:61}\\
  \Delta(c_i) &= c_i \otimes 1 + 1 \otimes c_i,\label{eq:79}\\
  \Delta(d_i) &= d_i \otimes 1 + 1 \otimes d_i. \label{eq:82}
\end{align}
where $\gamma_{(1)} = \gamma \otimes 1$, $\gamma_{(2)} = 1 \otimes
\gamma$.  In fact, the coproduct $\Delta$ is one of infinitely many
non-equivalent coproducts on the DIM algebra. In the literature on
refined topological strings the choice of coproduct is known as the
choice preferred direction (the coproducts are parametrized by
directions in the two-dimensional plane). For the sake of brevity we
will not explore here the issue of multiple coproducts, but simply
use Eqs.~\eqref{eq:60}--\eqref{eq:61} throughout the whole paper.

\paragraph{Fock representations.}
\label{sec:representations}
There exist representations of
$U_{q,t}(\widehat{\widehat{\mathfrak{gl}}}_1)$ on the Fock space
$\mathcal{F}(u)$ of a free chiral boson with momentum $u \in
\mathbb{C}^{\times}$ (also called spectral parameter). As we have
mentioned above, the representations of the algebra are organized into
$\widetilde{SL}(2,\mathbb{Z}) \times \mathfrak{S}_3$ orbits. The
family of Fock representations corresponds to the product of the orbit
of the vector $(1,0)$ under $SL(2,\mathbb{Z})$ action\footnote{The
  extra $\mathbb{Z}$ part of the universal cover can be absorbed by
  the shift of the spectral parameter $u$ and the rescaling of the
  generators.} and the orbit of a pair $(q,t^{-1})$ under permutations
of $(q,t^{-1},t/q)$. Hence, a Fock representation
$\mathcal{F}^{(r,s)}_{q,t^{-1}}(u)$ is characterized by a pair of
coprime integers $(r,s)$ and a choice of two out of three deformation
parameters.

The two examples of $\mathcal{F}^{(1,N)}_{q,t^{-1}}(u)$ ($N \in
\mathbb{Z}$) and $\mathcal{F}^{(0,1)}_{q,t^{-1}}(u)$ will be
particularly important for our investigation and we write them out
explicitly. The two representations are of course related by the
$\widetilde{SL}(2,\mathbb{Z})$ action, so one may wonder why consider
them separately. The reason for this is that we know the action of
$U_{q,t}(\widehat{\widehat{\mathfrak{gl}}}_1)$ on Fock representations
only in terms of the generating currents $x^{\pm}(z)$, $\psi^{\pm}(z)$
on which (unlike on $e_{(n,m)}$ for general $(n,m)$) the action of
$\widetilde{SL}(2,\mathbb{Z})$ is not explicit.

\begin{enumerate}
\item $\mathcal{F}^{(1,N)}_{q,t^{-1}}(u)$ usually called the
  \textbf{horizontal Fock representation}. The states of the Fock
  representation, which we denote $|\lambda,u\rangle$, are labelled by
  Young diagrams $\lambda$. They are in one to one correspondence with
  monomials of the form $a_{-\lambda_1} \cdots
  a_{-\lambda_n}|\varnothing,u\rangle$, where $a_n$, $n \neq 0$ are
  the creation and annihilation operators satisfying the commutation
  relations~\eqref{eq:86} and $|\varnothing,u\rangle$ is the vacuum
  state annihilated by $a_n$ with $n >0$.

The action of the generating currents in this representation
is~\cite{AFS}
\begin{align}
  \label{eq:63}x^{+}(z) &= \frac{u \left( z \sqrt{\frac{q}{t}} \right)^{-N}}{(1-q^{-1}) (1-t)} \exp \left[
    \sum_{n \geq 1} \frac{z^n}{n} (1-t^{-n}) a_{-n} \right] \exp
  \left[ - \sum_{n
      \geq 1} \frac{z^{-n}}{n} (1-t^n) a_n \right],\\
  x^{-}(z) &= \frac{u^{-1} \left( z \sqrt{\frac{q}{t}} \right)^N}{(1-q) (1-t^{-1})} \exp \left[ - \sum_{n
      \geq 1} \frac{z^n}{n} (1-t^{-n}) \left( \frac{t}{q}
    \right)^{\frac{n}{2}} a_{-n} \right] \exp \left[ \sum_{n \geq 1}
    \frac{z^{-n}}{n} (1-t^n) \left( \frac{t}{q}
    \right)^{\frac{n}{2}} a_n \right],\label{eq:64}\\
  \psi^{+}(z) &= \left( \frac{t}{q} \right)^{-\frac{N}{2}}\exp \left[ - \sum_{n\geq 1} \frac{z^{-n}}{n} (1-t^n)
    \left( 1 - \left( \frac{t}{q} \right)^n \right) \left( \frac{q}{t}
    \right)^{\frac{n}{4}} a_n
  \right],\label{eq:65}\\
  \psi^{-}(z) &= \left( \frac{t}{q} \right)^{\frac{N}{2}} \exp \left[ \sum_{n\geq 1} \frac{z^n}{n} (1-t^{-n})
    \left( 1 - \left( \frac{t}{q} \right)^n \right) \left( \frac{q}{t}
    \right)^{\frac{n}{4}} a_{-n} \right],\label{eq:66}\\
  \gamma &= \sqrt{\frac{t}{q}},\label{eq:67}\\
  K&= \left( \frac{t}{q} \right)^{\frac{N}{2}}.\label{eq:77}
\end{align}

\item $\mathcal{F}^{(0,1)}_{q,t^{-1}}(u)$ also known as the
  \textbf{vertical Fock representation}. It is convenient write down
  the action of the generating currents in this representation in the
  basis of Macodnald polynomials~\cite{MD-book}, i.e.\ assume that
  $|\lambda,u\rangle =
  P_{\lambda}^{(q,t)}(a_{-n})|\varnothing,u\rangle$:
  \begin{align}
    \label{eq:68}
    x^{+}(z) |\lambda,u\rangle &= \sum_{i=1}^{l(\lambda)+1}
    A^{+}_{\lambda,i} \delta \left( \frac{z}{u q^{\lambda_i} t^{1-i}}
    \right)
    |\lambda+1_i,u \rangle,\\
    x^{-}(z) |\lambda,u\rangle &= \sum_{i=1}^{l(\lambda)}
    A^{-}_{\lambda,i} \delta \left( \frac{z}{u q^{\lambda_i-1}
        t^{1-i}} \right)
    |\lambda-1_i,u \rangle, \label{eq:69}\\
    \psi^{+}(z) |\lambda,u\rangle &= \sqrt{\frac{q}{t}}\exp \left[
      \sum_{n \geq 1} \frac{1}{n} \left( \frac{u}{z} \right)^n \left(
        1 - (t/q)^n - \kappa_n
        \mathrm{Ch}_{\lambda} (q^n, t^{-n}) \right) \right] |\lambda,u\rangle,\label{eq:70}\\
    \psi^{-}(z) |\lambda,u\rangle &= \sqrt{\frac{t}{q}}\exp \left[
      \sum_{n \geq 1} \frac{1}{n} \left( \frac{z}{u} \right)^n \left(
        1 - (q/t)^n + \kappa_n \mathrm{Ch}_{\lambda} (q^{-n}, t^n)
      \right) \right] |\lambda,u\rangle, \label{eq:71}
\end{align}
where $\lambda \pm 1_i$ denotes the Young diagram with $\lambda_i$
replaced by $\lambda_i \pm 1$, $\mathrm{Ch}_{\lambda}(q,t)$ is defined
in Eq.~\eqref{eq:36} and
\begin{align}
  \label{eq:72}
  \kappa_n &=
  (1-q^n)(1-t^{-n})(1-(t/q)^n),\\
  A^{+}_{\lambda,i} &= \frac{1}{1-q^{-1}} \prod_{j=1}^i \psi \left(
    q^{\lambda_i - \lambda_j} t^{j-i} \right),\label{eq:74}\\
  A^{-}_{\lambda,i} &= - \frac{\sqrt{\frac{t}{q}}}{1-q} \frac{1 -
    q^{\lambda_i} }{1 - \frac{t}{q} q^{\lambda_i} }
  \prod_{j=i+1}^{l(\lambda)} \frac{\psi \left( q^{\lambda_i -
        \lambda_j-1} t^{j-i} \right)}{\psi \left( q^{\lambda_i - 1}
      t^{j-i} \right)},\label{eq:75}
\end{align}
and
\begin{equation}
  \psi(x) = \frac{(1 - t x) \left( 1 - \frac{q}{t} x\right)}{(1-x) (1
    - q x)}.\label{eq:76}
\end{equation}

\end{enumerate}

\paragraph{$qW_m$-reduction and $qq$-characters.}
\label{sec:qw_n-reduction}
Using the coproduct~\eqref{eq:60}--\eqref{eq:61} one defines the
action of the DIM algebra
$U_{q,t}(\widehat{\widehat{\mathfrak{gl}}}_1)$ on a tensor product of
$m$ horizontal Fock representations with different spectral parameters
$\bigotimes_{i=1}^m\mathcal{F}_{q,t^{-1}}^{(1,0)}(u_i)$.

The generators of the DIM algebra in this representation can be
expressed in terms of the currents of the smaller algebra: the
$qW_m$-algebra and an extra Heisenberg algebra. $qW_m$-algebra is
generated by currents of spins $s=2,3,\ldots,m$, which schematically
look as follows
\begin{equation}
  \label{eq:78}
  W^{(s)}(z) \sim \sum_{1\leq i_1 < i_2 < \ldots < i_{s-1}\leq m}u_{i_1} u_{i_2}
  \cdots u_{i_{s-1}} :\Lambda_{i_1}(z)
  \Lambda_{i_2}\left( \frac{q}{t} z \right)\cdots
  \Lambda_{i_{s-1}}\left( \left( \frac{q}{t} \right)^{s-2} z\right):,
\end{equation}
where $\Lambda_i(z)$ are certain free field vertex operators acting on
the tensor product of Fock representations. We denote the additional
Heisenberg current by $W^{(1)}(z)$.

The currents $W^{(s)}(z)$ are in fact the $qq$-characters~\cite{FR} of
$A_{m-1}$ type corresponding to fundamental representations $\Lambda^s
\mathbb{C}^m$. In $qq$-character approach the vertex operators
$\Lambda_i(z)$ are written in terms of the so-called
$\mathsf{Y}$-operators, certain elementary free boson exponents,
\begin{equation}
  \label{eq:81}
  \Lambda_i(z) =\, :\mathsf{Y}_i\left( \left( t/q \right)^{\frac{i-1}{2}} z \right) \mathsf{Y}_{i-1}\left(\left(t/q \right)^{\frac{i}{2}} z\right)^{-1}:,
\end{equation}
where one assumes that $\mathsf{Y}_0(z) = \mathsf{Y}_m(z) = 1$.

The action of the DIM current $x^{+}(z)$ on
$\bigotimes_{i=1}^m\mathcal{F}_{q,t^{-1}}^{(1,0)}(u_i)$ can be
expressed through the two lowest spin $W$-currents, the Hiesnberg
current $W^{(1)}(z)$ and $q$-deformed Virasoro stress-energy tensor
$W^{(2)}(z)$. Schematically we have (see~\cite{MMZ} for explicit
formulas)
\begin{multline}
  \label{eq:80}
  x^{+}(z)|_{\bigotimes_i \mathcal{F}(u_i)} \sim W^{(1)}(z) W^{(2)}(z) =
  W^{(1)}(z) \sum_{i=1}^m u_i  \Lambda_i(z) =\\
  =W^{(1)}(z) \sum_{i=1}^m
  u_i  :\mathsf{Y}_i\left( \left( t/q \right)^{\frac{i-1}{2}} z \right) \mathsf{Y}_{i-1}\left(\left(t/q \right)^{\frac{i}{2}} z\right)^{-1}:.
\end{multline}
The Heisenberg current $W^{(1)}(z)$ commutes with other $W$-currents,
and thus can be ignored in most formulas\footnote{Notice, how this
  property of the current $W^{(1)}(z)$ is similar to that of the
  ``empty crossing'' $\mathcal{T}^{\varnothing}_{\varnothing}$ from
  sec.~\ref{sec:stack-cross-top} given by Eq.~\eqref{eq:24}. Indeed,
  $\mathcal{T}^{\varnothing}_{\varnothing}$ depends on the same
  ``diagonal'' combination of free bosons $\sum_{i=1}^m
  (t/q)^{|n|(i-1)/2} a_n^{(i)}$ as the Heisenberg current. However, it
  includes only the positive modes, whereas $W^{(1)}(z)$ is a standard
  free field exponent depending on both positive and negative modes.}.

\section{Conventions and identities}
\label{sec:appendix}

In this Appendix we summarize some conventions and identities used in
the main text. The instanton counting partition function for (quiver)
$U(N)$ gauge theories with hypermultiplets in (anti)fundamental,
bi-fundamental and adjoint representations can be obtained using
standard Nekrasov formulas~\cite{Nekrasov:2002qd}. The partition
function is expressed as a sum over $N$-tuples of Young diagrams for
each node in the quiver while the summand is given by the products of
factors corresponding to each multiplet, which are built from the
following combinatorial factor $N_{\mu\nu}(Q)$:
\begin{align}
  \label{eq:14}\nonumber
  N_{\mu\nu}(Q)&=\prod_{(i,j)\in\mu}\left (1-Q\,q^{\mu_i-j}t^{\nu^{\mathrm{T}}_j-i+1} \right)\prod_{(i,j)\in\nu}\left ( 1-Q\,q^{-\nu_i+j-1}t^{-\mu^{\mathrm{T}}_{j}+i}\right)\\\nonumber
  &=\prod_{i,j=1}^{\infty}\frac{ (1-Q\,q^{\mu_{i}-j}t^{\nu^{\mathrm{T}}_{j}-i+1} )}{ (1-Q\,q^{-j}t^{-i+1} )}\\
  &=\prod_{i,j=1}^{\infty}\frac{(Q\,q^{\mu_{i}-\nu_{j}}t^{j-i+1};q)_{\infty}}{(Q\,q^{\mu_{i}-\nu_{j}}t^{j-i};q)_{\infty}}\frac{(Q\,t^{j-i};q)_{\infty}}{(Q\,t^{j-i+1};q)_{\infty}}.
\end{align}
There exist more representations of the factor $N_{\mu \nu}(Q)$, but
we will only use these three. It is evident from the first expression
in Eq.~\eqref{eq:14} that the factor is a polynomial. It also obeys
some nice identities:
\begin{equation}
  N_{\mu\nu}(Q)=(-v^{-1}Q)^{|\mu|+|\nu|}q^{\frac{\|\mu\|^2}{2}-\frac{\|\nu\|^2}{2}}t^{-\frac{\|\mu^t\|^2}{2}+\frac{\|\nu^t\|^2}{2}}N_{\mu\nu}(v^2Q^{-1}),
\end{equation}
where $|\mu|$ denotes the total number of boxes in the Young diagram
$\mu$, $\|\mu\|^2=\sum_{i=1}^{l(\mu)}\mu_i^2$ with $l(\mu)$ is the
number of non-empty rows of $\mu$, and $v$ is a shorthand for
$q^{1/2}t^{-1/2}$.

In our calculations, we have used an alternative expression for the
contributions of the $\mathsf{Y}$-operators which follows from the
identity
\begin{equation}
N_{\mu\square}(v^2 Q)=(1-Q)\frac{N_{\mu\varnothing}(t^{-1}Q) N_{\mu\varnothing}(qQ)}{N_{\mu\varnothing}(Q)}.\label{eq:122}
\end{equation}
The identity~\eqref{eq:122} can be proven using the representation of
the Nekrasov factor in terms of  $q$-Pochhammer symbols from the third
line of Eq.~\eqref{eq:14}.

These factors~\eqref{eq:122} have convenient degeneration properties,
and we make extensive use one of them:
\begin{equation}
N_{\mu\square}(1)=0,\qquad \text{unless}\quad \mu\in\{\varnothing,\square\}.
\end{equation}
Indeed, one can see that the boundary box in the first row,
$(1,\mu_1)\in\mu$, produces a vanishing contribution to the product in
$N_{\mu\square}(1)$ unless $\mu=\square$. Thus, to get a nonvanishing
answer for $N_{\mu\square}(1)$ the Young diagram $\mu$ should have the
form $[1,1,\ldots,1]$. Furthermore, a box with coordinates $(2,1)$ if
present also contributes a vanishing factor, so that the only
nontrivial possibility for $\mu$ is either $\square$ or $\varnothing$.
This property will allow us to truncate a sum over infinitely many
Young diagrams $\mu$ to a finite sum with only two terms.

The following useful identity can be proven by summing a geometric
progression:
\begin{equation}
  (1-q)\sum_{(i,j)\in\lambda}q^{j-1}t^{n-i+1}=\sum_{i=1}^{n}(1-q^{\lambda_i})t^{n+1-i},
  \label{eq:123}
\end{equation}
for any $n\geq l(\lambda)$, the length of the Young diagram
$\lambda$. Eq.~\eqref{eq:123} implies that
\begin{equation}
\label{powersum}
\frac{1}{1-q^{-n}}-(1-t^n)\mathrm{Ch}_{\lambda}(q^{-n},t^{-n})=-\frac{1-t^n}{1-q^n}t^{-n/2}q^n\,p_n(q^{-\lambda}t^{-\rho}),
\end{equation}
where $\mathrm{Ch}_{\lambda}$ is defined in Eq.~\eqref{eq:36} and
\begin{equation}
  \label{eq:124}
  p_n(q^{-\lambda} t^{-\rho}) = \sum_{i=1}^{\infty} q^{-n\lambda_i}
  t^{n \left( i - \frac{1}{2} \right)}.
\end{equation}

\section{Algebraic engineering of the $U(2)$ theory}
\label{sec:algebr-engin-su2}

In the main text we have introduced the algebraic approach to
calculating $qq$-characters based on the intertwining operators of the
DIM algebra $\mathcal{A}$. For the sake of completeness, we review in
this section the calculation of the instanton partition function of
the pure $U(2)$ theory \emph{without} spectator brane insertions using
similar technique which may be called algebraic
engineering. In~\cite{AFS}, the refined topological
vertices~\cite{RefTopVert} were shown to be matrix elements of
intertwining operators of the DIM algebra between triplets of Fock
spaces each labelled by a pair of integers and a spectral
parameter. From the Type IIB string view point, these integers are the
$(p,q)$-charges of 5-branes corresponding to Fock representations,
while in M-theory they encode the degeneration loci of a toric
diagram. The definitions of the Fock representations of the algebra
$\mathcal{A}$ are collected in Appendix~\ref{sec:basic-facts-about}.

In our calculations, we use the following intertwining operators
between triplets of Fock representations:
\begin{multline}
  \Psi^{\lambda}(x) =\quad \includegraphics[valign=c]{figures/top-vert-crop} \quad =
  (-ux)^{|\lambda|}(-x)^{-(N+1)|\lambda|}f_{\lambda}^{-(N+1)}\frac{q^{n(\lambda^{\mathrm{T}})}}{c_{\lambda}}\times\\
  \times \normord{
    \exp \left[ \sum_{n \neq 0} \frac{x^{-n}}{n} \left(
        \frac{1}{1-q^{-n}} - (1-t^n) \mathrm{Ch}_{\lambda}(q^{-n}
        t^{-n})\right) a_n\right]}=\\
  =(-ux)^{|\lambda|}(-x)^{-(N+1)|\lambda|}f_{\lambda}^{-(N+1)}\frac{q^{n(\lambda^{\mathrm{T}})}}{c_{\lambda}}\normord{\exp\left[
      -\sum_{n\neq 0} \frac{(q^{-1}
        t^{1/2}x)^{-n}}{n}\frac{1-t^n}{1-q^n}p_n(q^{-\lambda}t^{-\rho})\,a_n\right]}\label{eq:84}
\end{multline}
and its dual
\begin{multline}
  \Psi^{*}_{\mu}(y)=\quad \includegraphics[valign=c]{figures/top-vert-conj-crop} \quad =(-q^{-1}uy^{-1})^{-|\mu|}(-y)^{N|\mu|}f_{\mu}^{N}\frac{q^{n(\mu^{\mathrm{T}})}}{c_{\mu}}\times\\
  \times \normord{\exp
    \left[ -\sum_{n \neq 0} \frac{y^{-n}}{n} \left( \frac{t}{q}
      \right)^{\frac{|n|}{2}} \left( \frac{1}{1-q^{-n}} - (1-t^n)
        \mathrm{Ch}_{\mu}(q^{-n},
        t^{-n})\right) a_n\right]}=\\
  =(-q^{-1}uy^{-1})^{-|\mu|}(-y)^{N|\mu|}f_{\mu}^{N}\frac{q^{n(\mu^{\mathrm{T}})}}{c_{\mu}}\normord{\exp
    \left[ \sum_{n \neq 0}
      \frac{y^{-n}}{n}\frac{1-t^n}{1-q^n}q^{-|n|/2+n}t^{|n|/2-n/2}p_n(q^{-\mu}t^{-\rho})\,a_n\right]}\label{eq:85}
\end{multline}
where $\mathrm{Ch}_{\lambda}$, $n(\lambda^{\mathrm{T}})$,
$c_{\lambda}$ and $f_{\lambda}$ are defined by
Eqs.\eqref{eq:36},~\eqref{eq:113}--\eqref{eq:89} respectively, and we
have used the identity Eq.~\eqref{powersum}. Note that the
$(0,1)$-type Fock space $\mathcal{F}^{(0,1)}_{q,t^{-1}}$, whose states
are labelled by the diagrams $\lambda$ and $\mu$ in
Eqs.~\eqref{eq:84},~\eqref{eq:85}, is associated with the so-called
preferred direction of the refined topological vertex.

\begin{figure}[h]
  \begin{center}
    \includegraphics{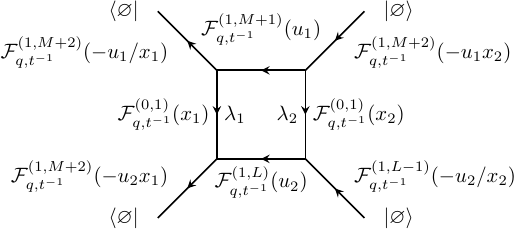}
  \end{center}
  \caption{Network of intertwining operators corresponding to the
    toric CY geometrically engineers $5d$ $U(2)$ pure gauge
    theory. The sum over the intermediate states $\lambda_1$,
    $\lambda_2$ is assumed.}
  \label{fig:SU2noR}
\end{figure}

The network giving rise to the $U(2)$ gauge theory is depicted in
Fig.~\ref{fig:SU2noR}. It consists of two intertwining intertwining
operators $\Psi$ and two dual intertwining operators $\Psi^{*}$.

There are two possible choices for the ``Chern-Simons level'' for the
$5d$ $U(2)$ theory, which is reflected in the choices for the levels
of the Fock representations in Fig.~\ref{fig:SU2noR}: $M=-1$ and $L=0$
corresponds to vanishing Chern-Simons level, and $M=-2$ and $L=0$
gives rise to level one, all other choices being either inconsistent
or equivalent to one of those two.

Generically, the sums over Young diagrams living on the edges along
the preferred directions (vertical in Fig.~\ref{fig:SU2noR}) cannot be
performed in a closed form for toric geometries with compact
four-cycles. Instead they correspond to the instanton sums in the
gauge theory that the toric diagram engineers. The same is true in the
algebraic approach as well and we will end up with the same instanton
sums over Young diagrams. In this Appendix we consider a quiver with a
single gauge node, hence a single set of vertical edges stretched
between two horizontal branes each associated with a Fock space. On
these Fock spaces act two sets of Heisenberg generators commuting with
each other:
\begin{gather}
  \left[a^{(1)}_n,a^{(1)}_m \right] = \left[a^{(2)}_n,a^{(2)}_m
  \right]
  = n \frac{1- q^{|n|}}{1 - t^{|n|}} \delta_{n+m,0},\\
  \left[a^{(1)}_n,a^{(2)}_m \right]=0, \qquad\text{for any}\quad
  (m,n)\in\mathbb{Z}^2,
\end{gather}
where the superscripts label the horizontal lines. Two such lines are
``glued'' to each other along a vertical edge by inserting a complete
basis of states in the vertical Fock representation, the basis of
Macdonald polynomials,
\begin{align}
\includegraphics[valign=c]{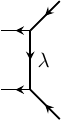}\quad =\quad   \sum_{\lambda}b_{\lambda}(q,t)\,\Psi^{*}_{\lambda}(x)
  \otimes\Psi^{\lambda}(x),
\end{align}
where $b_{\lambda}(q,t)$ is defined in Eq.~\eqref{eq:56}. The sums
over $\lambda_1$, $\lambda_2$ living on the vertical edges in
Fig.~\ref{fig:SU2noR} will turn into instanton sums in the gauge
theory partition function.

As depicted in Fig.~\ref{fig:SU2noR} we keep the external states in
the horizontal Fock spaces $\mathcal{F}^{(1,N)}$ empty, i.e.\ we
calculate the vacuum-to-vacuum matrix elements of the form $\langle
\varnothing|\otimes\langle \varnothing|\mathellipsis
|\varnothing\rangle \otimes |\varnothing\rangle$. We can proceed in
two different ways. We either insert complete sets of states between
all the intertwining operators in the picture and calculate matrix
elements of the individual operators before performing the sums over
intermediate states explicitly, or we construct the screening charges
by combining an intertwining operator $\Psi$ with the dual one
$\Psi^{*}$ coupled through a vertical edge. We choose the second
method and find that the partition function is given by the vacuum
matrix element of the following two operators
\begin{equation}
Z =  \sum_{\lambda_1, \lambda_2} \langle \varnothing|\otimes\langle \varnothing|
  \normord{\prod_{i\geq1}S(q^{-\frac{1}{2}}x_1q^{\lambda_{1,i}}t^{\frac{1}{2}
      -
      i})}\normord{\prod_{j\geq1}S(q^{-\frac{1}{2}}x_2q^{\lambda_{2,j}}t^{\frac{1}{2}
      - j})}|\varnothing\rangle
  \otimes |\varnothing\rangle,
\end{equation}
where the screening charge is
\begin{equation}
S(x)= \, \normord{\exp\left[ \sum_{n\neq 0}\frac{1}{n}\frac{1-t^n}{1-q^n}(1+q^n t^{-n})x^{-n}\alpha_n\right]}
\end{equation}
and $\alpha_n$ are deformed Heisenberg generators and can be expressed
as a linear combination of $a_n^{(1)}$, $a_n^{(2)}$ generators,
\begin{equation}
  \alpha_n\stackrel{\mathrm{def}}{=} \frac{1}{1+q^{|n|}t^{-|n|}}\left( a^{(1)}_{n}-q^{|n|/2}t^{-|n|/2} a^{(2)}_{n}\right).\label{eq:126}
\end{equation}

Using the Wick's theorem for the bosons~\eqref{eq:126} we get
\begin{multline}
  \normord{\prod_{i\geq1}S(q^{-1/2}x_1q^{\lambda_{1,i}}t^{1/2-i})}\normord{\prod_{j\geq1}S(q^{-1/2}x_2q^{\lambda_{2,j}}t^{1/2-j})}=N^{-1}_{\lambda_{2}\lambda_{1}}(x_2/x_1)N^{-1}_{\lambda_{2}\lambda_{1}}(v^2\,x_2/x_1)\times\\
  \times\normord{\prod_{i\geq1}S(q^{-1/2}x_1q^{\lambda_{1,i}}t^{1/2-i})\prod_{j\geq1}S(q^{-1/2}x_2q^{\lambda_{2,j}}t^{1/2-j})}.\label{eq:127}
\end{multline}
The normal ordered part in the right hand side of Eq.~\eqref{eq:127}
gives a simple factor when evaluated between the vacuum states. After
some straightforward manipulations, we get the following expression
for general $M$ and $L$:
\begin{multline}
Z=\sum_{\lambda_1,\lambda_2}\left( v^{-2}\frac{x_1 u_2}{x_2
    u_1}(-x_1)^{M-L+1}\right) ^{|\lambda_1|}\left( v^{-2}\frac{x_1
    u_2}{x_2 u_1}(-x_2)^{M-L+1}\right)
^{|\lambda_2|}f_{\lambda_1}^{M-L+1}f_{\lambda_2}^{M-L+1} \times\\
\times \left[ N_{\lambda_1\lambda_1}(1) N_{\lambda_2\lambda_2}(1)N_{\lambda_1\lambda_2}(x_1/x_2)N_{\lambda_2\lambda_1}(x_2/x_1)\right]^{-1},
\end{multline}
which reproduces the instanton partition function for $U(2)$ gauge
theory after certain identifications between the algebraic and gauge
theoretic variables.

\section{Elliptic deformation of vertex operators and DIM algebra}
\label{sec:elliptic-deformation}

Elliptic deformation of the DIM algebra as well as its free field
representation were introduced in~\cite{saito}. The deformation of the
algebra is written in terms of the generating currents~\eqref{eq:58},
\eqref{eq:59} with rational factors in their commutation relations
being replaced by Jacobi theta functions with elliptic modulus
$p=e^{\pi i \tau}$. The coproduct is formally given by the same
formulas~\eqref{eq:60}--\eqref{eq:82} as in the undeformed case.

In this Appendix we briefly review the elliptic deformation
prescription for free field vertex operators
(see~\cite{Ghoneim:2020sqi} for more details) which produces the
elliptic version of the \emph{horizontal} Fock representation of DIM
algebra starting from the formulas of
Appendix~\ref{sec:representations}. We also review the elliptic
deformation of the \emph{vertical} Fock representation introduced
in~\cite{Wang}.

\paragraph{Elliptic deformation of the horizontal Fock representation.}
\label{sec:ellipt-deform-horiz}

Suppose a vertex operator $X(z)$ has the following
generic mode expansion,
\begin{equation}
  X(z)=\exp\left(\sum_{n>0}X_{-n}^{-}z^n a_{-n} \right)\exp\left(\sum_{n>0}X_{n}^{+}z^{-n} a_{n} \right),\label{eq:106}
\end{equation}
where $a_n$ are the $(q,t)$-deformed Heisenberg generators satisfying
the commutation relations~\eqref{eq:86}. Notice that the generating
currents of DIM algebra in the horizontal Fock
representation~\eqref{eq:68}--\eqref{eq:71} also have the
form~\eqref{eq:106}.

It turns out that to reproduce elliptically deformed DIM commutation
relations it is not enough to deform the commutation
relations~\eqref{eq:86} of the generators $a_n$. One also has to
introduce another set of Heisenberg generators $b_n$ which commute
with the original generators and have slightly different commutation
relations among themselves. The elliptically deformed commutation
relations for the Heisenberg generators with elliptic parameter $p$
are as follows:
\begin{align}\nonumber
  &\left[a_n,a_m \right]=n(1-p^{|n|})\frac{1-q^{|n|}}{1-t^{|n|}}\delta_{m+n,0}, \qquad \left[ b_n,b_m\right]=n\frac{1-p^{|n|}}{(qt^{-1}p)^{|n|}}\frac{1-q^{|n|}}{1-t^{|n|}}\delta_{m+n,0}\\
  &\left[ a_n,b_m\right]=0.\label{eq:105}
\end{align}

The the elliptic version $X^{\mathrm{ell}}(z)$ of the vertex
operator~\eqref{eq:106} can be expressed as a product of two vertex
operators in terms of both sets of Heisenberg generators,
\begin{equation}
X^{\mathrm{ell}}(z)\stackrel{\mathrm{def}}{=} X^{(a)}(z)X^{(b)}(z),\label{eq:109}
\end{equation}
where
\begin{align}
  X^{(a)}(z) &\stackrel{\mathrm{def}}{=} \exp \left( \sum_{n>0} \frac{1}{1-p^n}X_{-n}^{-}z^n a_{-n}\right) \exp \left( \sum_{n>0} \frac{1}{1-p^n}X_{n}^{+}z^{-n} a_{n}\right),\label{eq:107}\\
  X^{(b)}(z) &\stackrel{\mathrm{def}}{=} \exp \left( -\sum_{n>0}
    \frac{p^n}{1-p^n}X_{n}^{-}z^{-n} b_{-n}\right) \exp \left(
    -\sum_{n>0} \frac{p^n}{1-p^n}X_{-n}^{+}z^{n}
    b_{n}\right).\label{eq:108}
\end{align}
Applying the prescription~\eqref{eq:109}--\eqref{eq:108} to DIM
algebra generators in the horizontal Fock
representation~\eqref{eq:63}--\eqref{eq:66} we obtain the horizontal
Fock representation of the elliptic DIM algebra.

\paragraph{Elliptic deformation of the vertical Fock representation.}
\label{sec:ellipt-deform-vert}
The vertical Fock representation~\eqref{eq:68}--\eqref{eq:71} can be
ellipticised without doubling the modes. The recipe for that is to
write elliptic Jacobi theta function
\begin{equation}
  \label{eq:112}
  \theta_p(x) = \prod_{k \geq 0}(1-p^{k+1})(1-p^kx) \left( 1 - \frac{p^{k+1}}{x} \right)
\end{equation}
in place of rational monomial $(1-x)$ in all the coefficient
functions:
  \begin{align}
    x^{+}_{\mathrm{ell}}(z) |\lambda,u\rangle &= \sum_{i=1}^{l(\lambda)+1}
    A^{+,\mathrm{ell}}_{\lambda,i} \delta \left( \frac{z}{u q^{\lambda_i} t^{1-i}}
    \right)
    |\lambda+1_i,u \rangle,\\
    x^{-}_{\mathrm{ell}}(z) |\lambda,u\rangle &= \sum_{i=1}^{l(\lambda)}
    A^{-,\mathrm{ell}}_{\lambda,i} \delta \left( \frac{z}{u q^{\lambda_i-1}
        t^{1-i}} \right)
    |\lambda-1_i,u \rangle, \\
    \psi^{+}_{\mathrm{ell}}(z) |\lambda,u\rangle &= \sqrt{\frac{q}{t}}\exp \left[
      \sum_{n \neq 0} \frac{1}{n} \frac{1}{1-p^{n}} \left( \frac{u}{z} \right)^n \left(
        1 - (t/q)^n - \kappa_n
        \mathrm{Ch}_{\lambda} (q^n, t^{-n}) \right) \right] |\lambda,u\rangle,\\
    \psi^{-}_{\mathrm{ell}}(z) |\lambda,u\rangle &= \sqrt{\frac{t}{q}}\exp \left[
      \sum_{n \neq 0} \frac{1}{n} \frac{1}{1-p^n}\left( \frac{z}{u} \right)^n \left(
        1 - (q/t)^n + \kappa_n \mathrm{Ch}_{\lambda} (q^{-n}, t^n)
      \right) \right] |\lambda,u\rangle, 
\end{align}
where
\begin{align}
  \label{eq:110}
  A^{+}_{\lambda,i} &= \frac{1}{\theta_p(q^{-1})} \prod_{j=1}^i
  \psi_{\mathrm{ell}} \left(
    q^{\lambda_i - \lambda_j} t^{j-i} \right),\\
  A^{-}_{\lambda,i} &= - \frac{\sqrt{\frac{t}{q}}\theta_p(q^{\lambda_i})}{\theta_p(q)\theta_p\left( \frac{t}{q}
      q^{\lambda_i}\right)} \prod_{j=i+1}^{l(\lambda)}
  \frac{\psi_{\mathrm{ell}} \left( q^{\lambda_i - \lambda_j-1} t^{j-i}
    \right)}{\psi_{\mathrm{ell}} \left( q^{\lambda_i - 1} t^{j-i}
    \right)},
\end{align}
and
\begin{equation}
  \label{eq:111}
  \psi_{\mathrm{ell}}(x) = \frac{\theta_p(t x) \theta_p\left( \frac{q}{t} x\right)}{\theta_p(x) \theta_p(q x)}.
\end{equation}

\end{document}